\documentclass[%
reprint,
superscriptaddress,
amsmath,amssymb,
aps,
prx,
]{revtex4-2}

\usepackage[utf8]{inputenc}
\usepackage[T1]{fontenc}
\usepackage{lmodern}
\usepackage[english]{babel}
\usepackage[autostyle=true]{csquotes}

\usepackage{graphicx}
\graphicspath{{figures/}}
\usepackage{bm}

\usepackage{hyperref}
\hypersetup{
	pdftitle={Non-invasive measurement of currents in analog quantum simulators},
	pdfauthor={Kevin T. Geier, Janika Reichstetter, Philipp Hauke},
}

\usepackage[dvipsnames]{xcolor}
\usepackage[separate-uncertainty=false]{siunitx}
\usepackage{mathtools}
\usepackage{dsfont}
\usepackage[caption=false]{subfig}

\usepackage[
capitalise,
]{cleveref}

\newcommand{\diff}{\mathop{}\!\mathrm{d}}
\newcommand{\etothepowerof}[1]{\mathrm{e}^{#1}}
\newcommand{\kb}{k_{\mathrm{B}}}
\newcommand{\commutator}[3][]{#1[ #2, #3 #1]}
\newcommand{\Commutator}[2]{\left[#1, #2\right]}
\newcommand{\anticommutator}[3][]{#1\{ #2, #3 #1\}}
\newcommand{\Anticommutator}[2]{\left\{#1, #2\right\}}

\DeclareMathOperator{\Tr}{Tr}
\newcommand{\doublestruck}[1]{\mathds{#1}}

\newcommand{\set}[1]{\{\,{#1}\,\}}

\newcommand{\ket}[1]{\mathinner{|{#1}\rangle}}
\newcommand{\braket}[2][]{\mathinner{#1\langle{#2}#1\rangle}}
\newcommand{\Braket}[1]{\mathinner{\left\langle{#1}\right\rangle}}
\newcommand{\ketbra}[2]{{|{#1}\rangle}{\langle{#2}|}}

\newcommand{\numancilla}{\ensuremath{n_{\mathrm{A}}}}
\newcommand{\pprojector}{\ensuremath{\mathcal{P}}}

\newcommand{\hcpl}{\ensuremath{\mathcal{H}_{\mathrm{cpl}}}}

\begin{document}

\title{Non-invasive measurement of currents in analog quantum simulators}

\author{Kevin T. Geier}
\thanks{These two authors contributed equally.\\E-mail K.T.G.\ at: \href{mailto:kevinthomas.geier@unitn.it}{kevinthomas.geier@unitn.it}\\E-mail J.R.\ at: \href{mailto:reichstetter@stud.uni-heidelberg.de}{reichstetter@stud.uni-heidelberg.de}}
\affiliation{INO-CNR BEC Center and Dipartimento di Fisica, Universit\`a di Trento, 38123 Povo, Italy}
\affiliation{Institute for Theoretical Physics, Ruprecht-Karls-Universität Heidelberg, Philosophenweg 16, 69120 Heidelberg, Germany}
\affiliation{Kirchhoff Institute for Physics, Ruprecht-Karls-Universität Heidelberg, Im Neuenheimer Feld 227, 69120 Heidelberg, Germany}

\author{Janika Reichstetter}
\thanks{These two authors contributed equally.\\E-mail K.T.G.\ at: \href{mailto:kevinthomas.geier@unitn.it}{kevinthomas.geier@unitn.it}\\E-mail J.R.\ at: \href{mailto:reichstetter@stud.uni-heidelberg.de}{reichstetter@stud.uni-heidelberg.de}}
\affiliation{Kirchhoff Institute for Physics, Ruprecht-Karls-Universität Heidelberg, Im Neuenheimer Feld 227, 69120 Heidelberg, Germany}
\affiliation{Institute for Theoretical Physics, Ruprecht-Karls-Universität Heidelberg, Philosophenweg 16, 69120 Heidelberg, Germany}

\author{Philipp Hauke}
\affiliation{INO-CNR BEC Center and Dipartimento di Fisica, Universit\`a di Trento, 38123 Povo, Italy}
\affiliation{Institute for Theoretical Physics, Ruprecht-Karls-Universität Heidelberg, Philosophenweg 16, 69120 Heidelberg, Germany}
\affiliation{Kirchhoff Institute for Physics, Ruprecht-Karls-Universität Heidelberg, Im Neuenheimer Feld 227, 69120 Heidelberg, Germany}

\date{\today}

\begin{abstract}
Despite the pristine abilities of analog quantum simulators to study quantum dynamics, possibilities to detect currents are sparse.
Here, we propose a flexible non-invasive technique to measure currents in quantum many-body systems by weakly coupling the system to an ancilla, followed by a measurement of the ancilla population.
We numerically benchmark the scheme at the example of interacting bosons in a Harper--Hofstadter optical-lattice ladder, and discuss potential experimental error sources.
The highly flexible protocol can be used with both hard-core and soft-core bosons as well as fermions, is easily extendable to more general observables like current--current correlations, and applies to other setups beyond cold atoms as we exemplify for the trapped-ion platform.
\end{abstract}

\maketitle

The investigation of quantum many-body systems in highly controllable quantum devices has accounted for major advances in understanding strongly correlated matter~\cite{Hauke2012,Cirac2012,Lewenstein2012,Bloch2012,Blatt2012,Schneider2012,Goldman2014,Chien2015,Gross2017,Galitski2019,Schaefer2020,Monroe2021}.
Such quantum simulators, e.g., based on cold atoms or trapped ions, offer the ability to observe phenomena as they evolve in real time and at a microscopic resolution, and they permit access to observables that are difficult to extract in solid-state samples.
However, one advantage of the solid state is the possibility to measure conduction properties by connecting wires to the sample~\cite{Imry1999,Datta2005}, which has enabled milestone discoveries such as the integer and fractional quantum Hall effects~\cite{Prange1990,Yoshioka2002,Goerbig2009}.
For ultracold atoms or trapped ions, such a coupling to the outside world would destroy the high-vacuum sample.
It is nevertheless possible to measure transport properties by microscopically tracking the evolution of the particle density~\cite{Fukuhara2013,Scherg2018,Brown2019,Nichols2019,Jepsen2020,Jurcevic2014,Smith2016,Maier2019}, by performing tomography after quenching an optical lattice~\cite{Kessler2014,Hauke2014,Flaeschner2016,Irsigler2019,Gluza2021}, or after dividing the sample into reservoir regions with different chemical potentials~\cite{Brantut2012,Krinner2015}.
In contrast, the direct measurement of currents requires additional experimental overhead, such as the coupling of a synthetic dimension to a cavity~\cite{Laflamme2017}.
Thus, it remains highly challenging to measure currents in quantum devices.

\begin{figure}
	\subfloat{\label{fig:measurementprotocol:ladder}}%
	\subfloat{\label{fig:measurementprotocol:couplingscheme}}%
	\subfloat{\label{fig:measurementprotocol:currentpatterns}}%
	\includegraphics[width=\columnwidth]{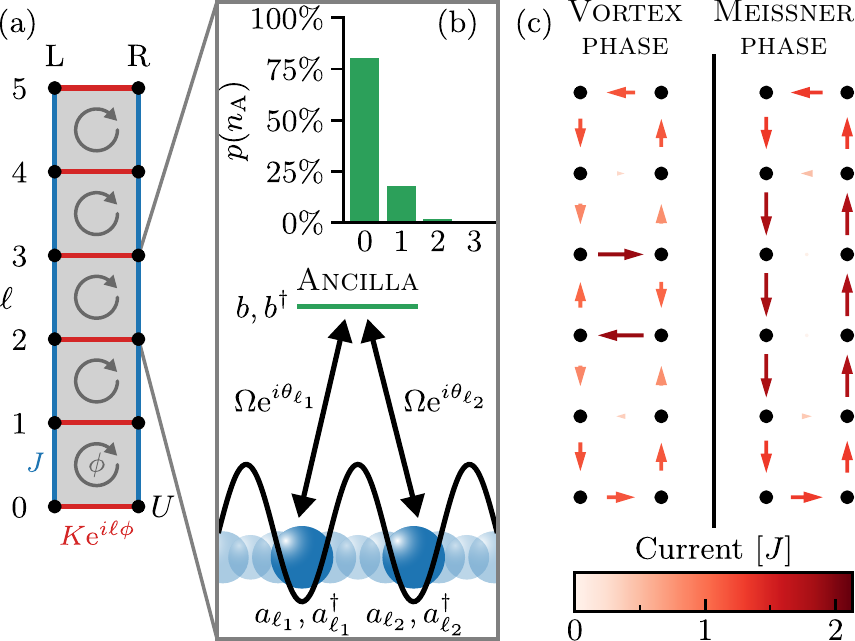}
	\caption{\label{fig:measurementprotocol}Schematic of the non-invasive measurement protocol for currents, illustrated for an optical lattice.
	(a)~We demonstrate the scheme at the example of bosons in a Harper--Hofstadter ladder with on-site interaction~$U$, real intra-leg tunneling~$J$, and complex inter-leg tunneling~$K \etothepowerof{i \ell \phi}$, generating a synthetic magnetic flux~$\phi$ per plaquette.
	(b)~A short pulse of strength~$\Omega$ coherently couples two sites $\ell_1$ and $\ell_2$ to an empty ancillary mode (see \protect\cref{eq:hamiltoniancoupling}). If the phases of the couplings are chosen appropriately (e.g., $\theta_{\ell_1} - \theta_{\ell_2} = \pi / 2$ for intra-leg currents), the current from site $\ell_1$ to $\ell_2$ can be extracted according to \protect\cref{eq:probability0} by measuring the probability of not populating the ancilla.
	(c)~The scheme reveals the current pattern of regularly-spaced vortices, characteristic for the vortex phase, as well as the chiral current running along the system boundary in the Meissner phase, shown here, respectively, for $K / J = \num{1.25}$ and $K / J = \num{2.5}$, as well as $\phi = 2 \pi / 3$ and $U / J = 1$.%
	}
\end{figure}

Here, we design a flexible and experimentally accessible protocol for measuring currents.
Our approach is based on the concept of non-invasive measurements~\cite{Wiseman2009,Svensson2013}, which allow one to access observables unobtainable via standard projective measurements, e.g., unequal-time correlations~\cite{Bednorz2013,Uhrich2017,Geier2021}.
The central idea is to weakly couple the system to an ancilla, on which suitable measurements are performed to extract information about the system.
Since the coupling is weak, measurement back action on the system is reduced and its state remains largely intact, though often at the price of a lower signal-to-noise ratio.
We show here, focusing in particular on optical lattice setups, that such a non-invasive scheme permits the accurate extraction of currents in quantum many-body systems.
The proposed protocol is illustrated in \cref{fig:measurementprotocol}: two lattice sites are coherently coupled to an ancilla with appropriately chosen phases, making the probability of populating the ancilla sensitive to the current between the coupled sites.
The scheme requires only the ability to distinguish an empty from a populated ancilla in the regime where the system's response to the coherent coupling is linear.
This requirement can readily be achieved, for instance, with modern quantum gas microscopes~\cite{Bakr2009,Sherson2010,Fukuhara2013,Nichols2019}, which also provide single-site addressing required to resolve local currents.

While the scheme works for bosons and fermions alike, both in and out of equilibrium, we benchmark our method using exact diagonalization at the example of equilibrium currents of interacting bosons in a Harper--Hofstadter ladder~\cite{Harper1955,Hofstadter1976,Bernevig2013,Aidelsburger2013,Miyake2013,Atala2014,Mancini2015,Stuhl2015,Tai2017,Dhar2012,Dhar2013,Huegel2014,Petrescu2015,Greschner2015,Greschner2016,Romen2018,Barbiero2020}.
This model mimics the Meissner effect in type-II superconductors exposed to an external magnetic field, and our technique directly reveals the characteristic current patterns of the Meissner and the vortex phases (see \cref{fig:measurementprotocol:currentpatterns}), as well as the transition to a Mott-insulating phase.
To further demonstrate the feasibility of the protocol, we discuss potential experimental sources of errors as well as strategies how to mitigate them.
Moreover, we propose possible extensions to other platforms, e.g., trapped ions, as well as to more general observables like current--current correlations.
Our approach thus opens the door to measuring fundamental conduction properties in a broad range of synthetic quantum systems realizing strongly correlated phases of matter.

\textit{Outline of the protocol.---}
We consider a general system described by the tight-binding Hamiltonian
\begin{equation}
	\label{eq:hamiltonian}
	\mathcal{H} = - \sum_{\ell \neq \ell^\prime} J_{\ell \ell^\prime} a_\ell^\dagger a_{\ell^\prime} + \mathcal{V} ,
\end{equation}
where $a_\ell$ ($a_\ell^\dagger$) denotes the bosonic or fermionic annihilation (creation) operator at local mode~$\ell$, which may represent lattice sites as well as internal states, and $\mathcal{V}$ contains any density--density interaction.
We allow for complex hopping amplitudes~$J_{\ell \ell^\prime} = J_{\ell^\prime \ell}^* = |J_{\ell \ell^\prime}| \etothepowerof{i \phi_{\ell \ell^\prime}}$ involving a Peierls phase~$\phi_{\ell \ell^\prime}$, as is common in models with synthetic gauge fields~\cite{Goldman2014,Galitski2019,Manovitz2020}.

It is our goal to measure expectation values involving the current operator from local mode $\ell_1$ to $\ell_2$,
\begin{equation}
\label{eq:currentoperator}
	j_{\ell_1 \ell_2} = -i \big( J_{\ell_1 \ell_2} a_{\ell_1}^\dagger a_{\ell_2} - J_{\ell_1 \ell_2}^* a_{\ell_2}^\dagger a_{\ell_1} \big) ,
\end{equation}
whose form follows by combining the Heisenberg equation of motion~$\partial_t n_{\ell_1} = i \commutator{\mathcal{H}}{n_{\ell_1}}$ with the continuity equation $\partial_t n_{\ell_1} + \vphantom{\sum_{\ell}}\smash{\sum_{\ell_2 \neq \ell_1}} j_{\ell_1 \ell_2} = 0$, expressing local conservation of the particle number (density)~$n_{\ell_1} = \vphantom{a^\dagger}\smash{a_{\ell_1}^\dagger} a_{\ell_1}$ (here and in what follows, we set $\hbar = 1$).

To formulate our non-invasive measurement protocol, we model the ancilla as a single bosonic or fermionic mode, which we assume to be initially empty.
We consider a coupling between system and ancilla according to the Hamiltonian
\begin{equation}
\label{eq:hamiltoniancoupling}
	\mathcal{H}^{\ell_1 \ell_2}_{\mathrm{cpl}} = \Omega \left( \etothepowerof{i \theta_{\ell_1}} b^\dagger a_{\ell_1} + \etothepowerof{i \theta_{\ell_2}} b^\dagger a_{\ell_2} \right) + \mathrm{h.c.} ,
\end{equation}
describing the $\Lambda$~configuration depicted in \cref{fig:measurementprotocol:couplingscheme}.
Here, the operator $b^\dagger$ ($b$) creates (annihilates) a particle in the ancilla, $\Omega$ is the coupling strength, and $\mathrm{h.c.}$ denotes the Hermitian conjugate.
This coupling scheme is guided by the intuition that pre-existing correlations between the modes $\ell_1$ and $\ell_2$ can be probed because they modify the interference between population transfer paths generated by the coherent coupling.
In order to access the current given by \cref{eq:currentoperator}, we choose the phases in \cref{eq:hamiltoniancoupling} such that $\theta_{\ell_2} - \theta_{\ell_1} = \phi_{\ell_1 \ell_2} - \pi / 2$ with $\phi_{\ell_1 \ell_2} = \arg(J_{\ell_1 \ell_2})$.
(Other choices, e.g., $\theta_{\ell_1} = \theta_{\ell_2}$, instead give access to the correlator $\vphantom{a^\dagger}\smash{a_{\ell_1}^\dagger} a_{\ell_2} + \smash{a_{\ell_2}^\dagger} a_{\ell_1}$, see \cref{app:derivation}.)
Experimentally, the ancilla can conveniently be realized as an additional internal level of the atoms, for which detuning or polarization of the optical lattice lasers are chosen such that it is trapped midway between the two sites under investigation. The coupling in \cref{eq:hamiltoniancoupling} between system and ancilla can then be realized by laser-assisted tunneling~\cite{Jaksch2003}.
Alternatively, the ancilla may also correspond to an offresonant site in an optical superlattice, where transitions between higher and lower sites can be generated via microwave pulses~\cite{SoltanPanahi2011}.

The coupling is applied as a short pulse of duration~$\Delta t$, whose shape is arbitrary as long as $\Delta t$ is much shorter than the characteristic time scales of the Hamiltonian~\labelcref{eq:hamiltonian}.
In this case, we can neglect the evolution under the system Hamiltonian~\labelcref{eq:hamiltonian} during the coupling, and the measurement can be regarded as taking an instantaneous snapshot of the system.
If $\rho = \rho_0 \otimes \ketbra{0}{0}$ is the combined state of system and ancilla before the coupling, the state after the coupling reads $\rho_{\ell_1 \ell_2}^\prime = U(\Delta t) \rho U^\dagger(\Delta t)$, where $U(\Delta t) = \exp (- i \smash{\mathcal{H}^{\ell_1 \ell_2}_{\mathrm{cpl}}} \Delta t)$ is the time evolution operator.
We now consider the probability of detecting $n_\mathrm{A}$ particles in the ancilla, given by $p_{\ell_1 \ell_2}(n_\mathrm{A}) = \Tr ( \rho_{\ell_1 \ell_2}^\prime \ketbra{n_\mathrm{A}}{n_\mathrm{A}} )$.
By expanding~$U(\Delta t)$ up to second order in $\Omega \Delta t$, we find that the probability of not detecting any particles in the ancilla is given by
\begin{equation}
\label{eq:probability0}
	p_{\ell_1 \ell_2}(0) = 1 - s \braket[\Big]{n_{\ell_1} + n_{\ell_2} + \frac{j_{\ell_1 \ell_2}}{|J_{\ell_1 \ell_2}|}} + \mathcal{O}(s^2) ,
\end{equation}
where we have introduced the effective coupling strength~$s = (\Omega \Delta t)^2$, and $\braket{\cdots}$ denotes the expectation value with respect to~$\rho_0$.
The derivation of this result is detailed in \cref{app:derivation}.
As \cref{eq:probability0} shows, the leakage of atoms out of the system is determined by the densities at the involved modes as well as by the current in between.
In the simplest case of a uniform system, the contribution of the densities to \cref{eq:probability0} merely yields a constant offset, while otherwise it can be accounted for via a separate standard measurement.
Alternatively, the combination $p_{\ell_1 \ell_2}(0) - p_{\ell_2 \ell_1}(0)$ is directly proportional to the current since $j_{\ell_1 \ell_2} = - j_{\ell_2 \ell_1}$ and the densities drop out.
The probability in \cref{eq:probability0}, i.e., the fraction of experimental runs where no atoms are found in the ancilla, may be extracted, e.g., with the help of a quantum gas microscope~\cite{Bakr2009,Sherson2010}, though other methods have been developed to resolve occupation probabilities for different sites of a superlattice~\cite{Yang2020}.
We stress that it is sufficient to be able to distinguish an empty ancilla from one with non-vanishing population, while resolving individual occupancies in the bosonic case can be used to enhance the accuracy of the current measurement (see below).

In a similar vein, the ability to resolve higher orders in the effective coupling~$s$ gives access to successively higher moments of the current operator~\labelcref{eq:currentoperator}.
For example, as shown in \cref{app:derivation:variances}, from the $s^2$ term in \cref{eq:probability0} it is possible to extract the expectation value of the combination ${(n_{\ell_1} + n_{\ell_2})^2} + {\anticommutator{n_{\ell_1} + n_{\ell_2}}{j_{\ell_1 \ell_2}}} / |J_{\ell_1 \ell_2}| + {j_{\ell_1 \ell_2}^2 / |J_{\ell_1 \ell_2}|^2}$.
To isolate the desired second moment of the current operator~$\braket{\vphantom{j_\ell}\smash{j_{\ell_1 \ell_2}^2}}$, the anti-commutator~$\braket{\anticommutator{n_{\ell_1} + n_{\ell_2}}{j_{\ell_1 \ell_2}}}$ can be obtained as a conditional expectation value of the densities following the coupling to the ancilla~\cite{Geier2021}, while the quantity~$\braket{(n_{\ell_1} + n_{\ell_2})^2}$ can be measured via standard density detection.
This enables access to the variance of the current, which has been used, for instance, to characterize the Mott-insulator--superfluid transition~\cite{Kessler2014}, as well as to reveal many-body multi-valued Lissajous figures~\cite{Metcalf2018} or transitions in the Aubry--André--Harper model~\cite{Roy2019}.

\begin{figure}
	\subfloat{\label{fig:benchmark:phasediagram}}%
	\subfloat{\label{fig:benchmark:chiralcurrent}}%
	\subfloat{\label{fig:benchmark:currentvariance}}%
	\subfloat{\label{fig:benchmark:ancillaprobability}}%
	\includegraphics[width=\columnwidth]{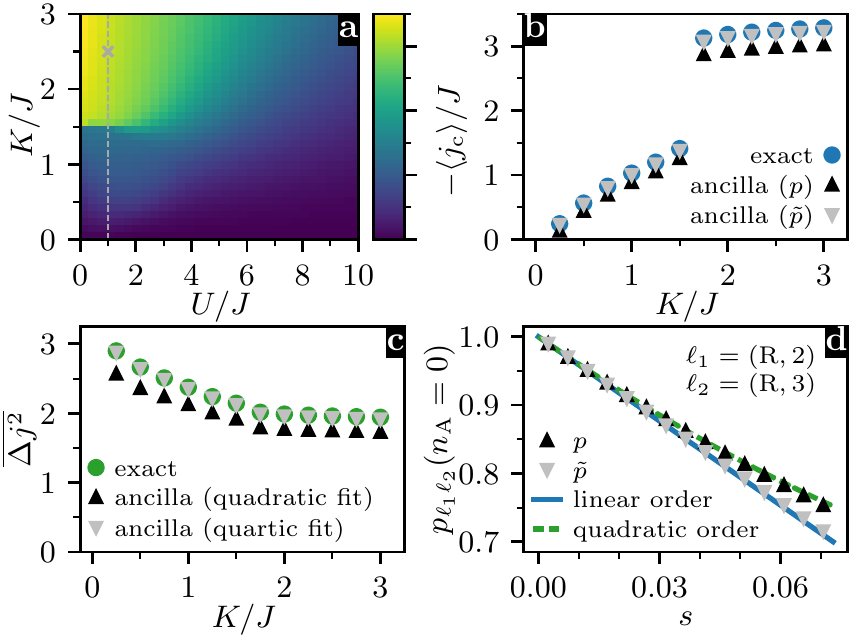}
	\caption{\label{fig:benchmark}Numerical benchmark of the current measurement scheme.
	(a)~Equilibrium phase diagram of interacting bosons in a Harper--Hofstadter ladder, computed using exact diagonalization for $N = 12$ particles at unit filling and a magnetic flux~$\phi = 2 \pi / 3$. At small on-site interactions~$U / J$, the system exhibits a transition from the vortex to the Meissner phase upon increasing the inter-leg tunneling strength~$K / J$, while at larger $U / J$, the system enters a Mott-insulating regime.
	(b)~Cross section of the chiral current~$\braket{j_\mathrm{c}}$ at $U / J = 1$ (grey dashed line in panel~a), in comparison with the values extracted from linear fits to the ancilla occupation probabilities $p(0)$ and $\tilde{p}(0) = 1 - [p(1) + 2 p(2)] / [1 - 2 s / 3]$.
	(c)~Mean current variance~$\overline{\Delta j^2}$ for $U / J = 1$, compared with a simulated measurement of this quantity involving quadratic and quartic fits to $p(0)$.
	(d)~Probability~$p_{\ell_1 \ell_2}(n_\mathrm{A} = 0)$ of not detecting any particles in the ancilla as a function of the effective coupling strength~$s$. The current is probed in the Meissner phase in positive flow direction for the same configuration as in \protect\cref{fig:measurementprotocol} ($K / J = \num{2.5}$ and $U / J = 1$, marked by the grey cross in panel~a).
	The continuous and dashed lines show the predictions from perturbation theory to linear and quadratic order in $s$, respectively. The linear regime for $\tilde{p}(0)$ is wider than for $p(0)$ due to the vanishing leading order error term, permitting a more accurate estimate of the current.%
	}
\end{figure}

\textit{Numerical benchmark.---}
We benchmark the proposed measurement scheme at the example of a Harper--Hofstadter model~\cite{Harper1955,Hofstadter1976,Bernevig2013}, which has successfully been realized experimentally in optical lattice setups~\cite{Aidelsburger2013,Miyake2013,Atala2014,Mancini2015,Stuhl2015,Tai2017}.
These systems and related variants host rich phase diagrams involving transitions between superfluid phases with persistent currents, Laughlin states, as well as (chiral) Mott-insulating phases~\cite{Dhar2012,Dhar2013,Huegel2014,Petrescu2015,Greschner2015,Greschner2016,Romen2018,Barbiero2020}.

Here, we focus on interacting bosons on a two-leg ladder in presence of an artificial magnetic field, as depicted in \cref{fig:measurementprotocol:ladder}.
Denoting sites as $\ell = (\ell_x, \ell_y)$ with $\ell_x \in \set{\mathrm{L}, \mathrm{R}}$ and $\ell_y \in \set{0, \dots, L - 1}$ labeling the ladder legs and rungs, respectively, this model is described by the Hamiltonian~\labelcref{eq:hamiltonian} with real hopping amplitudes along legs, $J_{(\ell_x,\ell_y),(\ell_x,\ell_y + 1)} \equiv J$, and complex hopping amplitudes across rungs, $J_{(\mathrm{L}, \ell_y),(\mathrm{R}, \ell_y)} \equiv K \smash{\etothepowerof{-i \phi \ell_y}}$. This way, each lattice plaquette is pierced by an effective magnetic flux~$\phi$.
Moreover, we consider on-site interactions at strength~$U$, i.e., $\mathcal{V} = U \sum_{\ell} n_\ell (n_\ell - 1) / 2$.
For our numerical benchmark, we use exact diagonalization on a ladder of length $L = 6$ with $N = 12$ particles (unit filling) and open boundary conditions. We focus on current detection across the ground state phase diagram, although our scheme is equally applicable to mixed states, e.g., at finite temperatures, as well as out of equilibrium.

To characterize the ground state phases, we use the chiral current $j_\mathrm{c} = j_\mathrm{L} - j_\mathrm{R}$ as an order parameter, where $j_{\ell_x} = \sum_{\ell_y = 0}^{L - 2} j_{(\ell_x, \ell_y), (\ell_x, \ell_y + 1)} / (L - 1)$ denotes the average current along the ladder leg~$\ell_x$~\cite{Atala2014}.
Its behavior across the phase diagram as resulting from our exact diagonalization is depicted in \cref{fig:benchmark:phasediagram}. 
At small $K / J$ and $U / J$, the model hosts a vortex phase, indicated by a small value of $\braket{j_\mathrm{c}}$ and currents circulating around plaquettes in regular distances (see \cref{fig:measurementprotocol:currentpatterns}).
Upon increasing $K / J$, the system undergoes a quantum phase transition to the Meissner phase, characterized by a large chiral current~$\braket{j_\mathrm{c}}$ along the ladder legs (see \cref{fig:measurementprotocol:currentpatterns}). This current mimics the expulsion of an external magnetic field, analogous to the Meissner effect in type-II superconductors.
Although the vortex--Meissner phase transition is continuous~\cite{Huegel2014}, the chiral current in our case exhibits a jump (see \cref{fig:benchmark:chiralcurrent}) due to finite-size effects.
At large $U / J$, the system enters a Mott-insulating regime, where any currents are suppressed.
As an additional benchmark observable, we use the mean current variance~$\overline{\Delta j^2} = N_{\mathrm{link}}^{-1} {\sum_{\braket{\ell, \ell^\prime}}} \Delta j_{\ell \ell^\prime}^2 / |J_{\ell \ell^\prime}|^2$, averaged over all $N_{\mathrm{link}} = 3 L - 2$ links between nearest neighbors, where $\Delta j_{\ell \ell^\prime}^2 = \braket{j_{\ell \ell^\prime}^2} - \braket{j_{\ell \ell^\prime}}^2$ is the variance of the current between the sites~$\ell$ and $\ell^\prime$.
As can be seen in \cref{fig:benchmark:currentvariance}, this quantity gradually decreases in the vortex phase as the tunneling~$K / J$ across the ladder rungs becomes stronger, until it saturates in the Meissner phase close to the value $\overline{\Delta j^2} = 2$ expected in the superfluid phase~\cite{Kessler2014}.

To simulate our measurement scheme, we compute the full evolution of system plus ancilla under the Hamiltonian $\mathcal{H} + \mathcal{H}_\mathrm{cpl}$ during a fixed coupling time~$J \Delta t = \num{0.01}$ for variable coupling strengths~$\Omega$.
In \cref{fig:benchmark:ancillaprobability}, we show the probability~$p_{\ell_1 \ell_2}(0)$ as a function of $s = (\Omega \Delta t)^2$, where the coupling is set up for probing a local leg current in the Meissner phase in positive flow direction (cf.~\cref{fig:measurementprotocol}).
For sufficiently small values of $s$, the result agrees well with the analytical approximations to linear and quadratic order in~$s$ (see \cref{eq:probability0} and \cref{app:derivation}), while higher-order non-linear effects become relevant as $s$ increases.
This behavior reflects the trade-off between maximizing the signal-to-noise ratio (large~$s$ preferred) and minimizing the systematic errors due to non-linearities (small~$s$ preferred).
For bosons, the linear regime can be significantly extended by resolving individual occupation numbers.
For instance, the combination~$\tilde{p}(0) = 1 - [p(1) + 2 p(2)] / [1 - 2 s / 3]$ agrees with \cref{eq:probability0} to linear order, while it is chosen such that the leading error term vanishes (see \cref{app:derivation:enhancements}).
As can be seen in \cref{fig:benchmark:ancillaprobability}, this quantity allows one to extract the linear slope more reliably, enabling a higher signal-to-noise ratio.

In \cref{fig:benchmark:chiralcurrent}, we compare the chiral current obtained from a simulation of the measurement scheme with the exact result.
To this end, we have extracted all constituent nearest-neighbor currents in positive flow direction, where the system's response to the coupling is stronger.
While in principle \cref{eq:probability0} allows for the extrapolation to $s = 0$ with arbitrary precision, in practice, a realistic signal-to-noise ratio requires a certain minimum fit range~$\Delta s$ (or equivalently $\Delta p$).
\Cref{fig:benchmark:chiralcurrent} shows the result for linear fits in the ranges $\Delta p = \SI{6}{\percent}$ and $\Delta \tilde{p} = \SI{20}{\percent}$ (the latter has also been used for \cref{fig:measurementprotocol:currentpatterns}), which represents a satisfactory compromise between accuracy and signal strength.
As expected, the chiral currents extracted from $\tilde{p}$ agree better with the exact result.

\Cref{fig:benchmark:currentvariance} shows a benchmark of the scheme for measuring the mean current variance~$\overline{\Delta j^2}$ (for details, see \cref{app:derivation:variances}). This quantity is obtained by extracting the local variance~$\Delta j_{\ell_1 \ell_2}^2$ of each nearest-neighbor current from the $s^2$~term of the probability~$p_{\ell_1 \ell_2} (0)$, which for bosons reads~$\partial^2 p_{\ell_1 \ell_2} (0) / \partial s^2 \big\rvert_{s = 0} = \braket{O_{\ell_1 \ell_2}^2} / 2 - \braket{O_{\ell_1 \ell_2}} / 3$ with $O_{\ell_1 \ell_2} = n_{\ell_1} + n_{\ell_2} + j_{\ell_1 \ell_2} / |J_{\ell_1 \ell_2}|$.
To isolate the quantity~$\braket{j_{\ell_1 \ell_2}^2}$, we assume that the surrounding terms have been obtained in a separate measurement, as discussed above.
Moreover, we probe the variance against the current's flow direction, as in this case the quadratic part of the probability is easier to resolve at small coupling strengths.
In \cref{fig:benchmark:currentvariance}, the variances have been extracted by fitting quadratic and quartic polynomials to the probabilities in the ranges~$\Delta p = \SI{6}{\percent}$ and $\Delta p = \SI{20}{\percent}$, respectively.
While the quadratic fits yield reasonable estimates of the variances, fitting higher-order polynomials can produce more accurate results, provided the quality of the data is sufficiently good.

Apart from the errors due to non-linearities, faulty detection of ancilla occupancies in the form of false positives or false negatives represents a principal experimental source of errors.
Let $\alpha$ and $\beta$ denote the rates of false positives and negatives, respectively (for simplicity, we do not distinguish different false negative probabilities). Then, instead of \cref{eq:probability0}, one obtains the modified result $p^\prime(0) = (1 - \alpha) p(0) + \beta p(n_\mathrm{A} > 0)$.
Assuming that $\alpha$ and $\beta$ do not depend on $s$, this leads to an irrelevant offset due to $\alpha$, but also to a modified slope by the factor $(1 - \alpha - \beta)$.
If estimates for $\alpha$ and $\beta$ are available, these systematic deviations can in principle be corrected for.

\textit{Discussion.---}
The proposed scheme lends itself to a variety of possible extensions. Further details to the following points can be found in~\cref{app:derivation,app:trappedions}

In many applications, e.g., for characterizing ground state phases as discussed above,
one is interested in global quantities like the chiral current. Although these can be deduced from several independent local measurements, it can be more efficient to simultaneously couple the relevant pairs of sites each to a distinct ancilla, e.g., the intermediate sites of an optical superlattice. The joint probability of not populating any ancilla then gives access to the sum of the individual local currents in a single measurement.
Furthermore, generalizations of the scheme give immediate access to loop currents around plaquettes~\cite{Goldbaum2008}, which characterize chiral insulators~\cite{Dhar2012,Dhar2013,Romen2018} and frustrated states of matter~\cite{GarciaRipoll2007}.
Assume for concreteness a system of ultracold bosons in a triangular optical lattice~\cite{Hauke2010,Hauke2013,Struck2013}.
By coupling the three sites of a plaquette simultaneously to a central ancilla, it is possible in certain scenarios to obtain the loop current with only two measurements.
Such measurements may help to detect the conjectured non-concomitant breaking of the $U(1)$ symmetry associated to magnetization and the $Z_2$ symmetry associated to the chirality of currents~\cite{Dhar2013}.

Our scheme also enables the measurement of current--current correlations in form of the anti-commutator $\braket{\anticommutator{j_{\ell_1 \ell_2}}{j_{\ell_3 \ell_4}}}$.
To this end, one couples pairs of sites $(\ell_1, \ell_2)$ and $(\ell_3, \ell_4)$ to a distinct ancilla each. The desired current--current correlation then appears in the $s^2$~term of the probability of not finding any particles in either ancilla.

Finally, although we have focused on applications in cold-atom systems, the scheme can equally well be applied to other platforms, simply by an appropriate choice of the ancillary level. For example, it enables the measurement of spin currents in trapped-ion quantum simulators of magnetic models~\cite{Blatt2012,Schneider2012,Monroe2021}. 
As detailed in \cref{app:trappedions}, the ancilla is represented by a collective vibrational mode of the ion crystal, e.g., the center-of-mass phonon mode, which may initially reside in a thermal mixed state. The ancilla can be coupled to the effective spin model under investigation via the red sideband transition, followed by a counting of the phonon population~\cite{Leibfried1996,Roos2000,Gebert2016,Um2016,Ding2017}.
Such a setup exploiting controlled spin--phonon coupling requires similar resources as those to investigate the Jaynes-- and Tavis--Cummings models~\cite{Pedernales2015,Retzker2007}, cooperative Jahn--Teller effects~\cite{Porras2012}, the spin-boson model~\cite{Porras2008,Lemmer2018}, quantum annealing~\cite{Nevado2016}, or lattice gauge theories~\cite{Yang2016,Davoudi2021}.

In conclusion, we have presented a versatile and accessible approach for the non-invasive measurement of current statistics.
Our numerical benchmarks at the example of a Harper--Hofstadter ladder demonstrate its potential for revealing current patterns across the entire phase diagram.
Promising targets for our protocol include currents in chiral Mott insulators~\cite{Dhar2012,Dhar2013,Romen2018} and fractional Hall states~\cite{Petrescu2015}, persistent currents in ring-shaped optical lattices~\cite{Kolovsky2006,Amico2014,Cominotti2015,Kohn2020}, as well as local Chern markers~\cite{Caio2019,Irsigler2019}.
Our protocol thus provides a blueprint for characterizing strongly correlated phases of matter in cold-atom-based quantum simulators and beyond.

\appendix

\begin{acknowledgments}
We thank M.\ Gärttner, S.\ Lannig, and M.\ K.\ Oberthaler for discussions,
and we acknowledge support by Provincia Autonoma di Trento and the ERC Starting Grant StrEnQTh (project ID $804305$).
This work is part of Q@TN --- Quantum Science and Technology in Trento --- and the Collaborative Research Centre ISOQUANT (project ID $273811115$).
The authors acknowledge support by the state of Baden-Württemberg through bwHPC.
\end{acknowledgments}

K.T.G.\ and J.R.\ contributed equally to this work.

\section{\label{app:derivation}Derivation of the non-invasive measurement protocol}

In this appendix, we discuss the mathematical framework underlying the non-invasive current measurement scheme presented in the main text.
Using time-dependent perturbation theory, we first derive the relevant formulas for the general case of an arbitrary number of system modes coupled to one or multiple ancillary modes.
We then specialize to the protocols discussed in the main text for measuring local and global currents, current variances, as well as current--current correlations.

\subsection{\label{app:derivation:general}General derivation}

We consider the general situation where an arbitrary number of system modes is coherently coupled to one or several of a total of $M$ bosonic or fermionic ancillary modes.
This scenario is described by the general coupling Hamiltonian
\begin{equation}
	\label{eq:hamiltoniancouplinggeneral}
	\hcpl = \sum_{m = 1}^{M} \Omega_m \left( b_m^\dagger A_m + A_m^\dagger b_m \right) ,
\end{equation}
where $\Omega_m \ge 0$ is the strength of the coupling to the $m$\nobreakdash-th ancilla with associated annihilation (creation) operators $b_m$ ($b_m^\dagger$).
The operator
\begin{equation}
	\label{eq:couplingoperator}
	A_m = \sum_{\ell} \lambda_{m \ell} a_\ell
\end{equation}
is a linear combination of system annihilation operators with (possibly complex) coefficients~$\lambda_{m \ell}$, specifying how the system mode~$\ell$ is coupled to the $m$\nobreakdash-th ancilla.
The $\Lambda$~configuration described by \cref{eq:hamiltoniancoupling} and depicted in \cref{fig:measurementprotocol:couplingscheme} is recovered for $M = 1$ if only two coefficients $\lambda_{1 \ell_1}$ and $\lambda_{1 \ell_2}$ are chosen different from zero.
For bosons, the annihilation and creation operators satisfy canonical commutation relations,
\begin{subequations}
	\label{eq:commutationrelations}
	\begin{align}
		\commutator[\big]{\alpha_i}{\beta_j} &= \commutator[\big]{\alpha_i^\dagger}{\beta_j^\dagger} = 0 , \\
		\commutator[\big]{\alpha_i}{\beta_j^\dagger} &= \delta_{\alpha \beta} \delta_{i j} ,
	\end{align}
\end{subequations}
while for fermions, they fulfill canonical anti-commutation relations,
\begin{subequations}
	\label{eq:anticommutationrelations}
	\begin{align}
		\anticommutator[\big]{\alpha_i}{\beta_j} &= \anticommutator[\big]{\alpha_i^\dagger}{\beta_j^\dagger} = 0 , \\
		\anticommutator[\big]{\alpha_i}{\beta_j^\dagger} &= \delta_{\alpha \beta} \delta_{i j} ,
	\end{align}
\end{subequations}
where $\alpha, \beta \in \{ a, b \}$ and $\delta_{ij}$ denotes the Kronecker delta.

Let $\rho_0$ denote the quantum state of the system under investigation.
This may be the ground state of a Hamiltonian, as considered in the numerical benchmarks in the main text,
a thermal state, or a state obtained after some unitary evolution.
We assume all ancillas to be initially in their respective vacuum states, such that when the coupling is turned on, the combined state of system and ancilla is given by the product state~$\rho = \rho_0 \otimes \ketbra{0 \cdots 0}{0 \cdots 0}$ (see \cref{app:trappedions:protocol} for generalizations to mixed initial states of the ancilla).
The coupling is applied as a short pulse of duration~$\Delta t$, which we assume to be sufficiently short compared to all other relevant time scales. It is then permissible to assume that the system evolves solely under the coupling Hamiltonian~\eqref{eq:hamiltoniancouplinggeneral} during the period~$\Delta t$.
Within these approximations, the coupling pulse can have an arbitrary shape~$f(t)$, normalized such that $\int_{0}^{\Delta t} \diff t \, f(t) = \Delta t$.
The time evolution during the coupling is governed by the von Neumann equation
\begin{equation}
	\dot{\rho}(t) = -i \Commutator{f(t) \hcpl}{\rho(t)} .
\end{equation}
Thus, the state of system plus ancilla after the coupling is given by
\begin{equation}
	\label{eq:stateaftercoupling}
	\rho^\prime \equiv \etothepowerof{-i \hcpl \Delta t} \rho \etothepowerof{i \hcpl \Delta t} .
\end{equation}

We are interested in the joint probability~$P(n_1, \dots, n_M)$ of finding $n_1$ particles in the first ancilla, $n_2$ particles in the second ancilla, and so on.
This probability can be expressed with the help of the projector $\pprojector_{n_1 \dots n_M} = \doublestruck{1} \otimes \ketbra{n_1 \cdots n_M}{n_1 \cdots n_M}$ as $P(n_1, \dots, n_M) = \Tr (\pprojector_{n_1 \dots n_M} \rho^\prime)$.
If the coupling strengths~$\Omega_m$ are sufficiently weak, we can expand the exponentials in \cref{eq:stateaftercoupling} to approximate this quantity perturbatively, yielding
\begin{widetext}
\begin{equation}
\begin{split}
	P(n_1, \dots, n_M) &= \delta_{n_1 0} \cdots \delta_{n_M 0} - \frac{1}{2} \Delta t^2 \Tr \left[ \pprojector_{n_1 \dots n_M} \left( \hcpl^2 \rho + \rho \hcpl^2 - 2 \hcpl \rho \hcpl \right) \right] \\
	&\phantom{={}} + \frac{1}{24} \Delta t^4 \Tr \left[ \pprojector_{n_1 \dots n_M} \left( \hcpl^4 \rho + \rho \hcpl^4 + 6 \hcpl^2 \rho \hcpl^2 - 4 \hcpl^3 \rho \hcpl - 4 \hcpl \rho \hcpl^3 \right) \right] + \mathcal{O}(\Delta t^6) .
\end{split}
\end{equation}
\end{widetext}
Note that since $\Tr (\pprojector_{n_1 \dots n_M} \hcpl^p \rho \hcpl^q) = 0$ if $p + q$ is odd, only even orders in $\Delta t$ contribute.
Up to quartic order in $\Delta t$, the probability that more than two particles are found in ancillary modes vanishes.
In what follows, we therefore focus on the probability~$P_0 = \Tr (\pprojector_{0 \dots 0} \rho^\prime)$ of not populating any ancilla, the probability~$P_1^{(m)} = \Tr (b_m^\dagger \pprojector_{0 \dots 0} b_m \rho^\prime)$ of finding a single particle in the $m$\nobreakdash-th ancilla (while all others are empty), the probability~$P_{2}^{(m_1, m_2)} = \Tr (b_{m_1}^\dagger b_{m_2}^\dagger \pprojector_{0 \dots 0} b_{m_2} b_{m_1} \rho^\prime)$ of detecting one particle in two distinct ancillas $m_1$ and $m_2$ each, and the probability~$P_2^{(m)} = \Tr [(b_m^\dagger)^2 \pprojector_{0 \dots 0} b_m^2 \rho^\prime] / 2$ of a double occupancy of the $m$\nobreakdash-th ancilla (which can be non-zero for bosons only).
After some algebra, using the (anti\nobreakdash-)commutation relations~\labelcref{eq:commutationrelations,eq:anticommutationrelations}, we find
\begin{widetext}
\begin{subequations}
\label{eq:probabilitiesgeneral}
\begin{align}
\label{eq:probabilitygeneral0}
	P_0 &= 1 - \sum_{m = 1}^{M} s_m \Braket{A_m^\dagger A_m}
	+ \sum_{m, k = 1}^{M} s_m s_k \Braket{ \frac{1}{3} A_m^\dagger A_m A_k^\dagger A_k + \frac{1}{6} A_m^\dagger A_k^\dagger A_k A_m } + \mathcal{O}(s^3) , \\
\label{eq:probabilitygeneral1}
	P_1^{(m)} &= s_m \Braket{A_m^\dagger A_m}
	- s_m \sum_{k = 1}^{M} s_k
	\Braket{\frac{1}{6} \anticommutator[\big]{A_m^\dagger A_m}{A_k^\dagger A_k}
	+ \frac{2}{3} A_m^\dagger A_k^\dagger A_k A_m} + \mathcal{O}(s^3) , \\
\label{eq:probabilitygeneral11}
	P_{2}^{(m_1, m_2)} &= s_{m_1} s_{m_2} \Braket{A_{m_1}^\dagger A_{m_2}^\dagger A_{m_2} A_{m_1}} + \mathcal{O}(s^3) , \\
\label{eq:probabilitygeneral2}
	P_2^{(m)} &= \frac{1}{2} s_m^2 \braket[\big]{\big( A_m^\dagger \big)^2 A_m^2} + \mathcal{O}(s^3) .
\end{align}
\end{subequations}
\end{widetext}
Here, $s_m = (\Omega_m \Delta t)^2$ and $\braket{\cdots} = \Tr (\cdots \rho_0)$ denotes the expectation value with respect to the system state~$\rho_0$.
Note that for fermions, the expectation values~$\braket{A_m^\dagger A_k^\dagger A_k A_m}$ vanish for $m = k$ due to the anti-commutation relations~\eqref{eq:anticommutationrelations}, in accordance with the Pauli exclusion principle.
It is easy to verify that the probabilities in \cref{eq:probabilitiesgeneral} correctly sum to unity,
\begin{equation}
P_0 + \sum_m P_1^{(m)} + \sum_{m_1 < m_2} P_{2}^{(m_1, m_2)} + \sum_m P_2^{(m)} = 1 + \mathcal{O}(s^3) .
\end{equation}

\subsection{\label{app:derivation:currents}Currents}

We now discuss how the general scheme derived in \cref{app:derivation:general} can be applied to measure currents.
In essence, this can be achieved by an appropriate choice of the operators~$A_m$ defined in \cref{eq:couplingoperator}.

In the most basic case, only one coefficient~$\lambda_{m \ell}$ is different from zero, i.e., only a single mode is coupled to one particular ancilla. Measuring the probabilities in \cref{eq:probabilitiesgeneral} then gives access to densities, their variances, as well as density--density correlations~\cite{Geier2021}. 

To access currents, at least two modes must be coupled to the same ancilla with appropriately chosen phases.
Consider the scenario of two modes~$\ell_1$ and $\ell_2$ coupled to a single ancilla ($M = 1$), as depicted in \cref{fig:measurementprotocol:couplingscheme}.
This setup is obtained from \cref{eq:couplingoperator} by setting $\lambda_{\ell_1} = \etothepowerof{i \theta_{\ell_1}}$, $\lambda_{\ell_2} = \etothepowerof{i \theta_{\ell_2}}$, and all others zero.
According to \cref{eq:probabilitygeneral0}, the probability~$p_{\ell_1 \ell_2}(0)$ of finding the ancilla empty is then given by
\begin{equation}
	\label{eq:probability0AdaggerA}
	p_{\ell_1 \ell_2}(0) = 1 - s \Braket{(A^\dagger A)_{\ell_1 \ell_2}} + \mathcal{O}(s^2)
\end{equation}
with
\begin{multline}
	\label{eq:AdaggerA}
	(A^\dagger A)_{\ell_1 \ell_2} = n_{\ell_1} + n_{\ell_2} \\
	+ \etothepowerof{i(\theta_{\ell_2} - \theta_{\ell_1})} a_{\ell_1}^\dagger a_{\ell_2} + \etothepowerof{-i(\theta_{\ell_2} - \theta_{\ell_1})} a_{\ell_2}^\dagger a_{\ell_1} .
\end{multline}
Choosing $\theta_{\ell_2} - \theta_{\ell_1} = \phi_{\ell_1 \ell_2} - \pi / 2$, we obtain $(A^\dagger A)_{\ell_1 \ell_2} = n_{\ell_1} + n_{\ell_2} + j_{\ell_1 \ell_2}$, where $\phi_{\ell_1 \ell_2} = \arg(J_{\ell_1 \ell_2})$ is the Peierls phase associated with the hopping amplitude~$J_{\ell_1 \ell_2}$ in the Hamiltonian~\eqref{eq:hamiltonian} and $j_{\ell_1 \ell_2}$ is the current operator defined in \cref{eq:currentoperator}.
\Cref{eq:probability0AdaggerA} then reduces to the result in \cref{eq:probability0}, giving access to the expectation value of the current~$\braket{j_{\ell_1 \ell_2}}$.

In fact, the freedom to adjust the phase difference $\theta_{\ell_2} - \theta_{\ell_1}$ enables access to both quadratures of the operator~$a_{\ell_1}^\dagger a_{\ell_2}$. For example, the choice~$\theta_{\ell_1} = \theta_{\ell_2}$ yields the correlator $\braket{a_{\ell_1}^\dagger a_{\ell_2} + a_{\ell_2}^\dagger a_{\ell_1}}$, which can be useful, e.g., for probing superfluidity~\cite{Pitaevskii2016}.

As discussed in the main text, the expectation values of the densities in \cref{eq:AdaggerA} can be obtained in a separate independent measurement, and their values can be subtracted from the measured probability~$p_{\ell_1 \ell_2}(0)$ to extract the expectation value of the current.
As an alternative, one can conduct an independent ancilla-based measurement of the probability~$p_{\ell_2 \ell_1}(0)$.
This exchange of the indices $\ell_1$ and $\ell_2$ does not affect the densities in \cref{eq:AdaggerA}, but reverses the sign of the current.
Thus, it is possible to extract the current according to
\begin{equation}
	\frac{\Braket{j_{\ell_1 \ell_2}}}{|J_{\ell_1 \ell_2}|} = \frac{\partial}{\partial s} \left[ \frac{p_{\ell_1 \ell_2}(0) - p_{\ell_2 \ell_1}(0)}{2} \right] \bigg\rvert_{s = 0} .
\end{equation}

Since to linear order in $s$ only single occupancies of the ancilla contribute, we have $p_{\ell_1 \ell_2}(1) = 1 - p_{\ell_1 \ell_2}(0)$, and therefore the same information as in \cref{eq:probability0AdaggerA} is contained in the expectation value of the ancilla population~$\braket{n_\mathrm{A}}_{\ell_1 \ell_2} = p_{\ell_1 \ell_2}(1) + \mathcal{O}(s^2)$.

Depending on the sign of the current, the system reacts more or less sensitive to the coherent coupling (cf.\ \cref{fig:fit:positive,fig:fit:negative}). This feature has the consequence that for positive currents, the effective perturbation strength~$s$ required to achieve a certain change of the ancilla occupation probability is much smaller than for negative currents. Nonetheless, the scheme yields comparable performance for measuring currents in either direction, allowing one to choose the configuration that best fits the experimental characteristics.

\subsection{\label{app:derivation:variances}Current variances}

Resolving the probabilities in \cref{eq:probabilitiesgeneral} to next-to-leading order gives access to higher moments involving the current operator, from which the variance of current can be extracted.
In what follows, we illustrate this possibility for the scenario of two modes $\ell_1$ and $\ell_2$ coupled to a single ancilla (see \cref{fig:measurementprotocol:couplingscheme}). We further choose the phases as discussed below \cref{eq:AdaggerA}, such that $(A^\dagger A)_{\ell_1 \ell_2} = n_{\ell_1} + n_{\ell_2} + j_{\ell_1 \ell_2} / |J_{\ell_1 \ell_2}|$ (from now on, we omit the subscripts $\ell_1$ and $\ell_2$ where there is no ambiguity).
For the special case of a single ancilla, the general expressions in \cref{eq:probabilitiesgeneral} simplify to
\begin{subequations}
	\label{eq:probabilitybosons}
\begin{align}
\label{eq:probabilitybosons0}
	p^{(\mathrm{B})}(0) &= 1 - s \Braket{A^\dagger A} + s^2 \Braket{\frac{1}{3} (A^\dagger A)^2 + \frac{1}{6} (A^\dagger)^2 A^2} , \\
\label{eq:probabilitybosons1}
	p^{(\mathrm{B})}(1) &= s \Braket{A^\dagger A} - s^2 \Braket{\frac{1}{3} (A^\dagger A)^2 + \frac{2}{3} (A^\dagger)^2 A^2} , \\
\label{eq:probabilitybosons2}
	p^{(\mathrm{B})}(2) &= \frac{1}{2} s^2 \Braket{ (A^\dagger)^2 A^2 } ,
\end{align}
\end{subequations}
for bosons, and
\begin{subequations}
	\label{eq:probabilityfermions}
\begin{align}
\label{eq:probabilityfermions0}
	p^{(\mathrm{F})}(0) &= 1 - s \Braket{A^\dagger A} + \frac{1}{3} s^2 \Braket{(A^\dagger A)^2} , \\
\label{eq:probabilityfermions1}
	p^{(\mathrm{F})}(1) &= s \Braket{A^\dagger A} - \frac{1}{3} s^2 \Braket{(A^\dagger A)^2} ,
\end{align}
\end{subequations}
for fermions.
(For conciseness of the formulas, here and in what follows we implicitly consider the expressions for the probabilities as perturbative approximations valid up to second order in~$s$, unless stated otherwise).

The relevant quantity for extracting the variance of the current~$\Delta j_{\ell_1 \ell_2}^2 = \braket{j_{\ell_1 \ell_2}^2} - \braket{j_{\ell_1 \ell_2}}^2$ is
\begin{equation}
	\label{eq:AdaggerA2}
\begin{split}
	(A^\dagger A)_{\ell_1 \ell_2}^2 &= \left( n_{\ell_1} + n_{\ell_2} \right)^2 + \frac{1}{|J_{\ell_1 \ell_2}|} \Anticommutator{n_{\ell_1} + n_{\ell_2}}{j_{\ell_1 \ell_2}} \\
	&\phantom{{}=} + \frac{j_{\ell_1 \ell_2}^2}{|J_{\ell_1 \ell_2}|^2} .
\end{split}
\end{equation}
For fermions, this combination is immediately accessible from \cref{eq:probabilityfermions0} as $\partial^2 p^{(\mathrm{F})}(0) / \partial s^2 \big\rvert_{s = 0}$. However, for bosons, the $s^2$~coefficient in \cref{eq:probabilitybosons0} contains an additional term~$\braket{(A^\dagger)^2 A^2}$.
This contribution can be accounted for via an independent measurement of one of the probabilities $p^{(\mathrm{B})}(1)$ or $p^{(\mathrm{B})}(2)$. It is then possible to eliminate the contribution due to $\braket{(A^\dagger)^2 A^2}$ by forming suitable linear combinations, e.g.,
\begin{equation}
p^{(\mathrm{B})}(0) - \frac{p^{(\mathrm{B})}(2)}{3} = 1 - s \Braket{A^\dagger A} + \frac{1}{3} s^2 \Braket{(A^\dagger A)^2} .
\end{equation}
Alternatively, \cref{eq:probabilitybosons0} can be simplified using the commutation relations~\eqref{eq:commutationrelations}.
For the coupling setup under consideration, we have $\commutator{A}{A^\dagger} = 2$, yielding
\begin{subequations}
	\label{eq:probabilitybosonscommuted}
\begin{align}
	p^{(\mathrm{B})}(0) &= 1 - s \Braket{A^\dagger A} + s^2 \Braket{\frac{1}{2} (A^\dagger A)^2 - \frac{1}{3} A^\dagger A} , \\
	p^{(\mathrm{B})}(1) &= s \Braket{A^\dagger A} - s^2 \Braket{(A^\dagger A)^2 - \frac{4}{3} A^\dagger A} , \\
	p^{(\mathrm{B})}(2) &= s^2 \Braket{\frac{1}{2} (A^\dagger A)^2 - A^\dagger A} .
\end{align}
\end{subequations}
The quantity~$\braket{(A^\dagger A)^2}$ can thus be obtained from the $s^2$~term of either of the above probabilites if the value of $\braket{A^\dagger A}$, corresponding to the linear coefficient of $p^{(\mathrm{B})}(0)$ or $p^{(\mathrm{B})}(1)$, is known.

In order to isolate the second moment of the current operator~$\braket{j_{\ell_1 \ell_2}^2}$ from \cref{eq:AdaggerA2}, knowledge of the other two terms is required. The density--density correlator $\braket{(n_{\ell_1} + n_{\ell_2})^2}$ is typically directly accessible, for instance, in quantum gas microscopes \cite{Bakr2009,Sherson2010}.
The contribution due to the density--current anti-commutator can be eliminated by conducting an additional ancilla-based measurement with the indices $\ell_1$ and $\ell_2$ exchanged. Since the densities in \cref{eq:AdaggerA2} are symmetric under this exchange, while the current is anti-symmetric, we have
\begin{equation}
	\label{eq:AdaggerA2sym}
	\frac{(A^\dagger A)_{\ell_1 \ell_2}^2 + (A^\dagger A)_{\ell_2 \ell_1}^2}{2} = \left( n_{\ell_1} + n_{\ell_2} \right)^2 + \frac{j_{\ell_1 \ell_2}^2}{|J_{\ell_1 \ell_2}|^2} .
\end{equation}
Alternatively, the quantity~$\anticommutator{n_{\ell_1} + n_{\ell_2}}{j_{\ell_1 \ell_2}}$ can be obtained as a conditional expectation value in the following way~\cite{Geier2021}.
Given that after the coupling $n_\mathrm{A}$ particles are detected in the ancilla, according to Lüders' rule~\cite{Lueders1950}, the state of the system conditioned on this measurement outcome reads
\begin{equation}
	\rho_{\numancilla}^\prime = \frac{\pprojector_{\numancilla} \rho^\prime \pprojector_{\numancilla}}{p(n_\mathrm{A})} ,
\end{equation}
where $\pprojector_{\numancilla} = \doublestruck{1} \otimes \ketbra{\numancilla}{\numancilla}$ is the projector on the subspace with $\numancilla$ particles in the ancilla, and $p(n_\mathrm{A}) = \Tr (\pprojector_{\numancilla} \rho^\prime)$ is the probability of detecting $\numancilla$ particles in the ancilla.
From \cref{eq:stateaftercoupling}, we find that the conditional state for $\numancilla = 0$, to leading order in $s$, is given by
\begin{equation}
	\rho_{\numancilla = 0}^\prime = \rho_0 - s \left( \frac{1}{2} \Anticommutator{A^\dagger A}{\rho_0} - \Braket{A^\dagger A} \rho_0 \right) ,
\end{equation}
where we have omitted the ancilla state.
A measurement of an observable~$O$ with respect to this state yields the conditional expectation value
\begin{equation}
	\Tr \left( O \rho_{\numancilla = 0}^\prime \right) = \Braket{O} - s \left( \frac{1}{2} \Braket{\Anticommutator{O}{A^\dagger A}} - \Braket{O} \Braket{A^\dagger A} \right) .
\end{equation}
Thus, measuring the observable $O = n_1 + n_2$, post-selected on the condition that no particle is detected in the ancilla, gives access to the desired density--current anti-commutator in \cref{eq:AdaggerA2}.

All in all, the variance of the current~$\Delta j_{\ell_1 \ell_2}^2 = \braket{j_{\ell_1 \ell_2}^2} - \braket{j_{\ell_1 \ell_2}}^2$ can be extracted by resolving the probabilities in \cref{eq:probabilitybosons,eq:probabilityfermions} to quadratic order in $s$, combined with suitable auxiliary measurements.

\begin{figure}
	\subfloat{\label{fig:fit:positive}}%
	\subfloat{\label{fig:fit:negative}}%
	\subfloat{\label{fig:fit:adaga}}%
	\subfloat{\label{fig:fit:current}}%
	\subfloat{\label{fig:fit:adaga2}}%
	\subfloat{\label{fig:fit:current2}}%
	\includegraphics[width=\columnwidth]{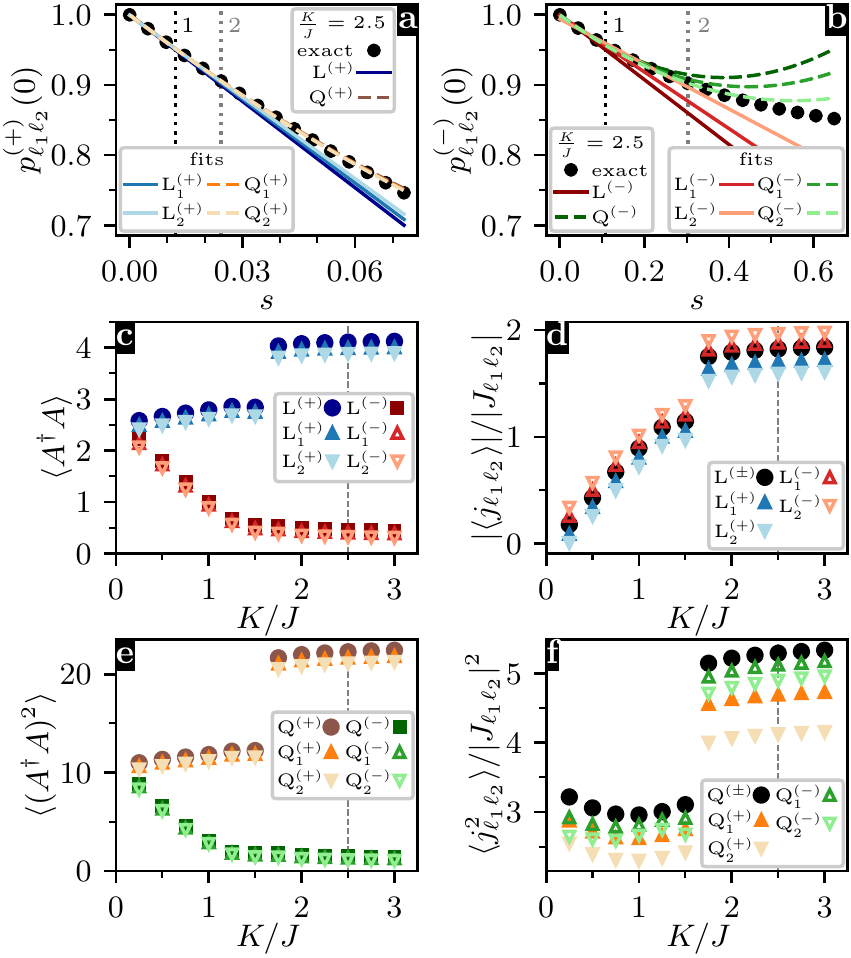}
	\caption{\label{fig:fit}Performance of the scheme for measuring currents and their variances in positive and negative flow direction.
		(a, b)~Probability~$p_{\ell_1 \ell_2}^{(\pm)}$ of not detecting any particles in the ancilla as a function of the effective coupling strength~$s = (\Omega \Delta t)^2$ for a bosonic ladder with rung hopping strength~$K / J = \num{2.5}$ (dashed vertical lines in central and lower panels), on-site interaction~$U / J = 1$, and magnetic flux~$\phi = 2 \pi / 3$.
		The lattice sites $\ell_1 = (\mathrm{R}, 2)$ and $\ell_2 = (\mathrm{R}, 3)$ are coupled such that the current is probed in positive~($+$) or negative~($-$) flow direction.
		The numerically computed probabilities are compared to the analytical predictions in \protect\cref{eq:probabilitybosonscommuted} to linear~($\mathrm{L}$) and quadratic~($\mathrm{Q}$) order in~$s$.
		To extract the first and second moment of the current operator, we fit, respectively, linear and quadratic polynomials to~$p_{\ell_1 \ell_2}^{(\pm)}$ in the ranges $\Delta p = \SI{5}{\percent}$~($1$) and $\Delta p = \SI{10}{\percent}$~($2$), as marked by the dotted vertical lines.
		For negative currents~(b), the system responds less strong to the coupling, making the linear and quadratic orders in $s$ easier to distinguish than for positive currents (a).
		(c)~The expectation value of the operator~$A^\dagger A = n_{\ell_1} + n_{\ell_2} + j_{\ell_1 \ell_2} / |J_{\ell_1 \ell_2}|$ extracted from the linear fits~($\mathrm{L}_1^{(\pm)}$ and $\mathrm{L}_2^{(\pm)}$) tends to underestimate the exact result~($\mathrm{L}^{(\pm)}$).
		(d)~Currents computed from the values of $\braket{A^\dagger A}$ in (c). The magnitudes of the currents are systematically underestimated (overestimated) for measurements in positive (negative) flow direction.
		(e, f)~While the values of $\braket{(A^\dagger A)^2}$~(e) extracted from the quadratic fits~($\mathrm{Q}_1^{(\pm)}$ and $\mathrm{Q}_2^{(\pm)}$) deviate only little from the exact result~($\mathrm{Q}^{(\pm)}$), the derived second moments of the current operator~$\braket{j_{\ell_1 \ell_2}^2}$~(f) exhibit large errors for those values based on $p^{(+)}$. Consequently, $p^{(-)}$ is preferred for measuring current variances.%
	}
\end{figure}

As shown in \cref{fig:fit}, the performance of the scheme varies depending on the direction in which the variance is probed.
As for positive currents the quantity~$\braket{(A^\dagger A)^2}$ is larger (see \cref{fig:fit:adaga2}), this quantity needs to be extracted more accurately to get the desired expectation value~$\braket{j_{\ell_1 \ell_2}^2}$ with a given precision than it is the case for negative currents.
In addition, since the linear coefficient~$\braket{A^\dagger A}$ is smaller for negative currents (see \cref{fig:fit:adaga}), the quadratic regime is easier to resolve when probing the current in this direction (cf.\ \cref{fig:fit:positive,fig:fit:negative}).
For these reasons, it is preferable to probe the variance against the flow direction of the current (see \cref{fig:fit:current2}).%

\subsection{\label{app:derivation:enhancements}Enhancing the signal-to-noise ratio}

An inherent difficulty typical of non-invasive measurement protocols is the low signal-to-noise ratio.
This is because these protocols rely on the assumption that the coupling between system and ancilla is weak.
As a consequence, the formulas in \cref{eq:probabilitiesgeneral} for extracting currents are valid only in the linear regime.
Thus, it can be challenging for an experiment, on the one hand, to make the coupling sufficiently weak to access the linear regime, and, on the other hand, to obtain a reasonably strong signal.

Increasing the coupling strength beyond the linear regime leads to a systematic deviation of the measured values.
As can be seen in \cref{eq:probabilitygeneral0,eq:probabilitygeneral1}, the linear and the quadratic terms have opposite signs.
Thus, the magnitude of the linear slope is typically underestimated in a linear fit, which leads to a systematic underestimation (overestimation) of the magnitudes of positive (negative) currents, as exemplified in \cref{fig:fit:current}. The accuracy of the measurement may therefore be improved by combining measurements in positive and negative flow direction.
For bosons, the range of the linear regime can in general self-consistently be assessed using the condition that at most a single particle is detected in the ancilla, as higher occupancies can only stem from non-linear processes.

Nonetheless, it is possible to mitigate the error due to non-linearities using knowledge about higher occupation probabilites.
This works, naturally, only for bosons, as for fermions, $p(0)$ and $p(1)$ contain the same information due to the Pauli exclusion principle.

To eliminate the leading order error term in \cref{eq:probability0AdaggerA}, we exploit the fact that the quantities $\braket{(A^\dagger A)^2}$ and $\braket{A^\dagger A}$ appear with different coefficients in the $s^2$ terms of the probabilities~\eqref{eq:probabilitybosonscommuted}.
This can be achieved by considering the combinations $p(1) + 2 p(2)$ or $2 - 2 p(0) - p(1)$, yielding
\begin{equation}
	\frac{p(1) + 2 p(2)}{1 - 2 s / 3} = s \Braket{A^\dagger A} + \mathcal{O}(s^3) .
\end{equation}
As shown in \cref{fig:benchmark:ancillaprobability}, this quantity exhibits a significantly extended linear regime, allowing one to operate at higher signal-to-noise ratios, which ultimately leads to a more accurate measurement.

\subsection{Current--current correlations}

Measuring the probabilities in \cref{eq:probabilitiesgeneral} to quadratic order in $s$ gives access to current--current correlations.
In what follows, we discuss this possibility for correlations~$\braket{j_{\ell_1 \ell_2} j_{\ell_3 \ell_4}}$ of the current operators $j_{\ell_1 \ell_2}$ and $j_{\ell_3 \ell_4}$ between two pairs of modes $(\ell_1, \ell_2)$ and $(\ell_3, \ell_4)$, each coupled to a different ancilla.
The coupling operators in \cref{eq:couplingoperator} then read $A_{1, \ell_1 \ell_2} = \etothepowerof{i \theta_{1 \ell_1}} a_{\ell_1} + \etothepowerof{i \theta_{1 \ell_2}} a_{\ell_2}$ and $A_{2, \ell_3 \ell_4} = \etothepowerof{i \theta_{2 \ell_3}} a_{\ell_3} + \etothepowerof{i \theta_{2 \ell_4}} a_{\ell_4}$, respectively.
For each pair of modes, the phases~$\theta_{m \ell}$ are chosen such that $A_1^\dagger A_1 = n_{\ell_1} + n_{\ell_2} + j_{\ell_1 \ell_2} / |J_{\ell_1 \ell_2}|$ and $A_2^\dagger A_2 = n_{\ell_3} + n_{\ell_4} + j_{\ell_3 \ell_4} / |J_{\ell_3 \ell_4}|$ (cf.~\cref{app:derivation:currents}).

We first discuss the case where the modes~$\ell_j$ with $j = 1, \dots, 4$ are all distinct.
Then, the operators $A_1$ and $A_2$ (anti\nobreakdash-)commute for bosons (fermions), and \cref{eq:probabilitygeneral0} simplifies to
\begin{multline}
	p_{\ell_1 \ell_2, \ell_3 \ell_4}(0, 0) = 1 + \left[ p_{\ell_1 \ell_2}(0) - 1 \right] \\
	+ \left[ p_{\ell_3 \ell_4}(0) - 1 \right] + s_1 s_2 \braket[\big]{A_1^\dagger A_1 A_2^\dagger A_2} .
\end{multline}
Here, the probabilities $p_{\ell \ell^\prime}(0)$ are given by \cref{eq:probabilitybosons0} or \cref{eq:probabilityfermions0}, obtained by coupling a pair of modes $(\ell, \ell^\prime)$ to a single ancilla.
The cross term
\begin{multline}
	\label{eq:twositecorrelatorsimple}
	A_1^\dagger A_1 A_2^\dagger A_2 = \left( n_{\ell_1} + n_{\ell_2} \right) \left( n_{\ell_3} + n_{\ell_4} \right) + \frac{j_{\ell_1 \ell_2} j_{\ell_3 \ell_4}}{|J_{\ell_1 \ell_2} J_{\ell_3 \ell_4}|} \\
	+ \frac{1}{|J_{\ell_3 \ell_4}|} \left( n_{\ell_1} + n_{\ell_2} \right) j_{\ell_3 \ell_4} + \frac{1}{|J_{\ell_1 \ell_2}|} j_{\ell_1 \ell_2} \left( n_{\ell_3} + n_{\ell_4} \right)
\end{multline}
contains the desired current--current correlator.
To isolate it, one can pursue similar strategies as in \cref{app:derivation:variances}.
That is, one can measure the surrounding density--density correlations and density--current correlations independently, where the latter can be obtained as conditional expectation values.
As an alternative, due to the symmetries of the densities and the currents with respect to exchanging the indices, the combination
\begin{multline}
	\frac{1}{4} \left[ p_{\ell_1 \ell_2, \ell_3 \ell_4} - p_{\ell_2 \ell_1, \ell_3 \ell_4} - p_{\ell_1 \ell_2, \ell_4 \ell_3} + p_{\ell_2 \ell_1, \ell_4 \ell_3} \right] (0, 0) \\
	= s_1 s_2 \Braket{j_{\ell_1 \ell_2} j_{\ell_3 \ell_4}}
\end{multline}
gives direct access to the current--current correlator.

If not all coupled modes are distinct, the procedure is more involved due to the non-commutativity of the associated operators.
We elucidate this circumstance for current--current correlations between two adjacent sites, corresponding to $\ell_2 = \ell_3$ and $\ell_1 \neq \ell_4$.
Then, applying the (anti\nobreakdash-)commutation relations
$\commutator{A_1}{A_2^\dagger} = \etothepowerof{i(\theta_{1 \ell_2} - \theta_{2 \ell_2})}$ ($\anticommutator{A_1}{A_2^\dagger} = \etothepowerof{i(\theta_{1 \ell_2} - \theta_{2 \ell_2})}$) for bosons (fermions) to \cref{eq:probabilitygeneral0}, we find
\begin{multline}
	p_{\ell_1 \ell_2, \ell_2 \ell_4}(0, 0) = 1 + \left[ p_{\ell_1 \ell_2}(0) - 1 \right] + \left[ p_{\ell_2 \ell_4}(0) - 1 \right] \\
	+ s_1 s_2 \left( \frac{1}{2} \braket[\big]{ \anticommutator[\big]{A_1^\dagger A_1}{A_2^\dagger A_2} } - \frac{1}{6} R_{\ell_1 \ell_2, \ell_2 \ell_4} \right) .
\end{multline}
with
\begin{equation}
\begin{split}
	R_{\ell_1 \ell_2, \ell_2 \ell_4} &= \braket[\big]{ \etothepowerof{i(\theta_{1 \ell_2} - \theta_{2 \ell_2})} A_1^\dagger A_2
	+ \mathrm{h.c.}} \\
	&= \frac{\Braket{j_{\ell_1 \ell_2}}}{|J_{\ell_1 \ell_2}|} + 2 \Braket{n_{\ell_2}} + \frac{\Braket{j_{\ell_2 \ell_4}}}{|J_{\ell_2 \ell_4}|} \\
	&\phantom{{}=} - \braket[\big]{ \etothepowerof{i(\phi_{\ell_1 \ell_2} + \phi_{\ell_2 \ell_4})} a_{\ell_1}^\dagger a_{\ell_4}
	+ \mathrm{h.c.} } .
\end{split}
\end{equation}
The last term in $R_{\ell_1 \ell_2, \ell_2 \ell_4}$ can in principle be obtained by directly coupling the sites $\ell_1$ and $\ell_4$ to a single ancilla with appropriately chosen phases.
However, it may be more practicable to eliminate this contribution all together by considering instead the combination
\begin{equation}
\begin{split}
	&p_{\ell_1 \ell_2, \ell_3 \ell_4}(0, 0) - \frac{1}{3} p_{\ell_1 \ell_2, \ell_3 \ell_4}(1, 1) \\
	&\quad = 1 + \left[ p_{\ell_1 \ell_2}(0) - 1 \right] + \left[ p_{\ell_3 \ell_4}(0) - 1 \right] \\
	&\quad \phantom{{}= 1} + \frac{1}{3} s_1 s_2 \braket[\big]{\anticommutator[\big]{A_1^\dagger A_1}{A_2^\dagger A_2}} .
\end{split}
\end{equation}
The last term
\begin{multline}
	\anticommutator[\big]{A_1^\dagger A_1}{A_2^\dagger A_2} = \Anticommutator{n_{\ell_1} + n_{\ell_2}}{n_{\ell_2} + n_{\ell_4}} + \frac{\Anticommutator{j_{\ell_1 \ell_2}}{j_{\ell_2 \ell_4}}}{|J_{\ell_1 \ell_2} J_{\ell_2 \ell_4}|} \\
	+ \frac{1}{|J_{\ell_2 \ell_4}|} \Anticommutator{n_{\ell_1} + n_{\ell_2}}{j_{\ell_2 \ell_4}} + \frac{1}{|J_{\ell_1 \ell_2}|} \Anticommutator{n_{\ell_2} + n_{\ell_4}}{j_{\ell_1 \ell_2}}
\end{multline}
then contains the desired current--current correlator~$\braket{\anticommutator{j_{\ell_1 \ell_2}}{j_{\ell_2 \ell_4}}}$, which can be isolated in a similar way as discussed before.
Note that compared to \cref{eq:twositecorrelatorsimple}, the anti-commutator appears here since the operators $j_{\ell_1 \ell_2}$ and $j_{\ell_2 \ell_4}$ do not commute.
The ability to simultaneously measure observables that are incompatible according to the Heisenberg uncertainty principle is a typical feature of non-invasive measurement protocols.

\subsection{\label{app:derivation:loopcurrents}Loop currents}

\begin{figure}
	\includegraphics[width=\columnwidth]{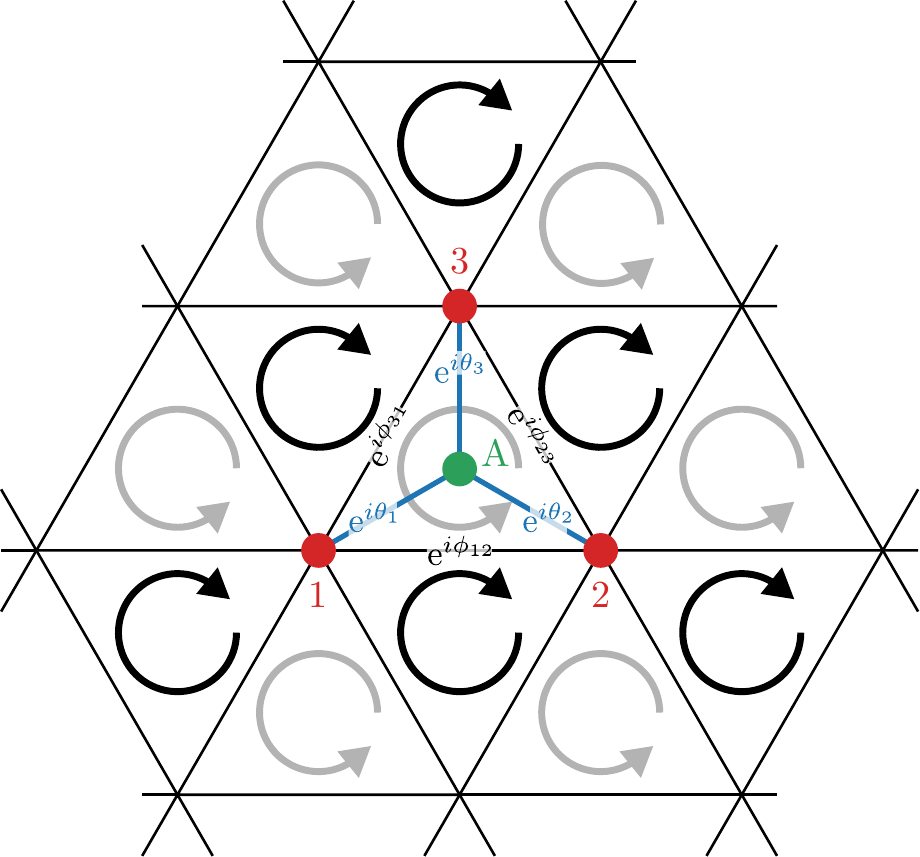}
	\caption{\label{fig:loopcurrent}Coupling scheme for the measurement of loop currents.
		Under geometric frustration, cold atoms in optical lattices can assume patterns of chiral symmetry breaking with persistent loop currents (indicated by the circular arrows). By coupling a given plaquette, spanned by the sites $1$, $2$, and $3$, to a central ancilla ($\mathrm{A}$), it is possible in certain scenarios to extract the loop current with only two measurements.%
	}
\end{figure}

The current measurement scheme can also be applied to detect persistent loop currents around lattice plaquettes, which arise, for example, as a consequence of chiral symmetry breaking in frustrated systems~\cite{GarciaRipoll2007,Hauke2010,Hauke2013,Struck2013}.
By coupling all sites spanning a plaquette simultaneously to the same ancilla, the number of measurements required to obtain the loop current can in some cases be reduced in comparison to separate measurements of the involved nearest-neighbor currents.

We illustrate this possibility for the triangular lattice depicted in \cref{fig:loopcurrent}.
The goal is to extract the expectation value of the loop current operator around a plaquette,
\begin{equation}
	\label{eq:loopcurrent}
	j_\triangle = j_{12} + j_{23} + j_{31} .
\end{equation}

To this end, the three lattice sites forming a plaquette are coupled to a central ancilla via the coupling operator
\begin{equation}
	A = r_1 \etothepowerof{i \theta_1} a_1 + r_2 \etothepowerof{i \theta_2} a_2 + r_3 \etothepowerof{i \theta_3} a_3 .
\end{equation}
Here, we choose the magnitudes of the couplings according to
\begin{equation}
	r_1 = \sqrt{\frac{\zeta_{12} \zeta_{31}}{\zeta_{23}}} , \quad
	r_2 = \sqrt{\frac{\zeta_{12} \zeta_{23}}{\zeta_{31}}} , \quad
	r_3 = \sqrt{\frac{\zeta_{23} \zeta_{31}}{\zeta_{12}}} ,
\end{equation}
where $\zeta_{\ell \ell^\prime} = |J_{\ell \ell^\prime}| / J$ denotes the magnitude of the hopping amplitude~$J_{\ell \ell^\prime}$, relative to some arbitrary energy scale~$J > 0$.
To obtain the correct Peierls phases~$\phi_{\ell \ell^\prime} = \arg(J_{\ell \ell^\prime})$ of the currents, it is desirable to choose the phases of the couplings as
\begin{subequations}
\begin{align}
	\theta_{2} - \theta_{1} &= \phi_{12} - \alpha \pmod{2 \pi} , \\
	\theta_{3} - \theta_{2} &= \phi_{23} - \beta \pmod{2 \pi} , \\
	\theta_{1} - \theta_{3} &= \phi_{31} - \gamma \pmod{2 \pi} .
\end{align}
\end{subequations}
However, by summing these equations, one can see that the angles $\alpha$, $\beta$, and $\gamma$ cannot be chosen arbitrarily, but they must satisfy the constraint
\begin{equation}
	\label{eq:constraint}
	\alpha + \beta + \gamma \pmod{2 \pi} = \Phi ,
\end{equation}
where $\Phi = \phi_{12} + \phi_{23} + \phi_{31} \pmod{2 \pi}$ is the effective magnetic flux through the plaquette.
We then obtain
\begin{equation}
	\label{eq:AdaggerAplaquette}
\begin{split}
	A^\dagger A &= r_1^2 n_1 + r_2^2 n_2 + r_3^2 n_3 \\
	&\phantom{{}=} + \frac{1}{J} \left[ \cos (\alpha) c_{12} + \cos (\beta) c_{23} + \cos (\gamma) c_{31} \right] \\
	&\phantom{{}=} + \frac{1}{J} \left[ \sin (\alpha) j_{12} + \sin (\beta) j_{23} + \sin (\gamma) j_{31} \right] ,
\end{split}
\end{equation}
where $c_{\ell \ell^\prime} = J_{\ell \ell^\prime} a_{\ell}^\dagger a_{\ell^\prime} + J_{\ell \ell^\prime}^* a_{\ell^\prime}^\dagger a_{\ell}$ denotes the correlator between the sites $\ell$ and $\ell^\prime$.

For a magnetic flux~$\Phi = \pm \pi / 2$, one can choose $\alpha = \beta = \gamma = \mp \pi / 2$, such that all correlators vanish and \cref{eq:AdaggerAplaquette} directly gives access to the loop current~\eqref{eq:loopcurrent}.
Another important special case is the fully frustrated configuration with $\Phi = \pm \pi$. Then, one measurement with $\alpha = \beta = \gamma = \pm \pi / 3$ yields $A^\dagger A = n_\triangle + c_\triangle / 2 J \pm \sqrt{3} j_\triangle / 2 J$, and a second one with $\alpha = \beta = \gamma = \pm \pi$ gives $A^\dagger A = n_\triangle - c_\triangle / J$, where $c_\triangle = c_{12} + c_{23} + c_{31}$ and $n_\triangle = r_1^2 n_1 + r_2^2 n_2 + r_3^2 n_3$. From these two measurements, and possibly an additional standard measurement of the densities, one can readily extract~$j_\triangle$.
This constitutes an advantage over individual measurements of the involved nearest-neighbor currents, where at least three ancilla-based measurements (plus additional standard measurements of the density) are required.

For general phases~$\phi_{\ell \ell^\prime}$, consistent choices of $\alpha$, $\beta$, and $\gamma$ are not possible due to the constraint~\eqref{eq:constraint}, such that additional measurements can become necessary to isolate the desired loop current.
If one is, however, interested in currents in the laboratory frame, the relevant current operator
\begin{equation}
	\label{eq:loopcurrentlabframe}
	j_{\ell \ell^\prime}^{\mathrm{(lab)}} = -i |J_{\ell \ell^\prime}| \big( a_{\ell_1}^\dagger a_{\ell_2} - a_{\ell_2}^\dagger a_{\ell_1} \big) ,
\end{equation}
does not involve Peierls phases.
This situation corresponds to a measurement with $\Phi = 0$, which permits the choices $\alpha = \beta = \gamma = 2 \pi / 3$, yielding $A^\dagger A = n_\triangle - c_\triangle^{\mathrm{(lab)}} / 2 J + \sqrt{3} j_\triangle^{\mathrm{(lab)}} / 2 J$, and $\alpha = \beta = \gamma = 0$, giving $A^\dagger A = n_\triangle + c_\triangle^{\mathrm{(lab)}} / J$.
This combination allows the extraction of the loop current~\eqref{eq:loopcurrentlabframe} with only two ancilla-based measurements, independent of the effective magnetic flux in the co-moving frame.

\subsection{Global current statistics}

So far, we have focused the discussion on local currents such as nearest-neighbor or loop currents, as well as current--current correlations involving two pairs of modes.
Beyond these basic building blocks, the scheme can immediately be extended to more general observables involving multiple local currents, e.g., currents through or into a given lattice site.
In some situations, one is even interested in global currents like the chiral current used in the main text to characterize the ground state phases of bosons in a Harper--Hofstadter ladder.
Although global currents can be calculated from a summation of local ones, it can be more efficient to measure the global quantity of interest directly.
Furthermore, in setups without single-site addressing, only global quantities are typically accessible.

Within the framework of our non-invasive measurement protocol, the sum of arbitrary local currents can be obtained directly by simultaneously coupling the relevant pairs of modes each to a distinct ancilla, located, for instance, at the intermediate sites of an optical superlattice.
According to \cref{eq:probabilitygeneral0}, to linear order in $s$, the probability of not populating any ancilla then gives access to the desired sum of local currents, while the corresponding variance can be extracted from the quadratic order in $s$.

In what follows, we discuss this scenario for a measurement of the chiral current in a Harper--Hofstadter ladder (see main text).
We consider a total of $M = 2 L$ ancillas located midway between the system lattice sites on the ladder legs. They are labeled by the index $m = (m_x, m_y)$ with $m_x \in \set{\mathrm{L}, \mathrm{R}}$ and $m_y \in \set{0, \dots, {L - 1}} + 1 / 2$, where we consider periodic boundary conditions for ease of notation. The ancilla $(m_x = \ell_x, m_y = l_y + 1 / 2)$ is then coupled to the lattice sites $(l_x, l_y)$ and $(l_x, l_y + 1)$, and the phases are chosen (see \cref{app:derivation:currents}) such that
\begin{subequations}
\begin{gather}
\begin{split}
	(A^\dagger A)_{(\mathrm{L}, \ell_y + 1 / 2)} &= n_{(\mathrm{L}, \ell_y)} + n_{(\mathrm{L}, \ell_y + 1)} \\
	&\phantom{{}=} + \frac{1}{J} j_{(\mathrm{L}, \ell_y), (\mathrm{L}, \ell_y + 1)} ,
\end{split} \\
\begin{split}
(A^\dagger A)_{(\mathrm{R}, \ell_y + 1 / 2)} &= n_{(\mathrm{R}, \ell_y)} + n_{(\mathrm{R}, \ell_y + 1)} \\
&\phantom{{}=} - \frac{1}{J} j_{(\mathrm{R}, \ell_y), (\mathrm{R}, \ell_y + 1)} .
\end{split}
\end{gather}
\end{subequations}
For simplicity, we assume the hopping amplitudes along the ladder legs to be of equal magnitude~$J \equiv |J_{(l_x, l_y), (l_x, l_y \pm 1)}|$, although spatial anisotropies in the hopping amplitudes can be accounted for by adjusting the relative magnitudes of the coefficients~$\lambda_{m \ell}$ in \cref{eq:couplingoperator} appropriately (cf.\ \cref{app:derivation:loopcurrents}.
This configuration then yields
\begin{equation}
	\sum_{m} A_m^\dagger A_m = \frac{L}{J} j_\mathrm{c} + 2 N ,
\end{equation}
where the chiral current operator for periodic boundary conditions reads
\begin{equation}
	j_\mathrm{c} = \frac{1}{L} \sum_{l_y = 0}^{L - 1} \left[ j_{(\mathrm{L}, \ell_y), (\mathrm{L}, \ell_y + 1)} - j_{(\mathrm{R}, \ell_y), (\mathrm{R}, \ell_y + 1)} \right] ,
\end{equation}
and $N = \sum_{\ell} n_{\ell}$ is the total particle number operator, which reduces to a constant when working in a subspace with a fixed number of particles.
The probability of not finding any particles in any ancilla~\eqref{eq:probabilitygeneral0} thus becomes
\begin{equation}
	P_0 = 1 - s \Braket{\frac{L}{J} j_{\mathrm{c}} + 2 N} + \mathcal{O}(s^2)
\end{equation}
with $s \equiv s_m = (\Omega_m \Delta t)^2$, giving access to the chiral current~$\braket{j_\mathrm{c}}$.

By resolving this probability up to quadratic order in~$s$, it is possible to also obtain the variance of the chiral current $\Delta j_\mathrm{c}^2 = \braket{j_\mathrm{c}^2} - \braket{j_\mathrm{c}}^2$.
To eliminate the terms $\braket{A_m^\dagger A_k^\dagger A_k A_m}$ in \cref{eq:probabilitygeneral0}, it is convenient to consider the quantity $P_0 - P_2 / 3$, where
\begin{equation}
P_2 = \frac{1}{2} \sum_{m_1 \neq m_2} P_2^{(m_1, m_2)} + \sum_{m} P_2^{(m)}
\end{equation}
is the probability of finding two particles in ancillary modes all together.
We then obtain, up to quadratic order in~$s$, the result
\begin{equation}
\begin{split}
P_0 - \frac{1}{3} P_2 &= 1 - s \sum_m \Braket{A_m^\dagger A_m} + \frac{1}{3} s^2 \braket[\Big]{ \Big( \sum_m A_m^\dagger A_m \Big)^2 } \\
&= 1 - s \Braket{\frac{L}{J} j_{\mathrm{c}} + 2 N} + \frac{1}{3} s^2 \braket[\bigg]{ \bigg( \frac{L}{J} j_{\mathrm{c}} + 2 N \bigg)^2 } ,
\end{split}
\end{equation}
from which the variance of the chiral current can be extracted in a similar way as described in \cref{app:derivation:variances}.
Such a global measurement using multiple ancillas can be much more efficient than measuring the constituent local currents and pairwise current--current correlations individually.

\section{\label{app:trappedions}Trapped-ion implementation}

In this appendix, we discuss how to implement our non-invasive current measurement protocol in trapped-ion platforms.
To this end, we first specify the class of Hamiltonians as well as the type of currents we intend to investigate.
We then present a possible implementation of the measurement scheme, where a collective vibrational mode plays the role of the ancilla.
Unlike in our previous discussion, where we assumed the ancilla to be empty, we consider the ancillary collective mode to be thermally occupied, which is a common scenario in trapped-ion systems.
We present a generalization of the measurement scheme adapted to this setup, and discuss how to harness standard tools of trapped-ion experiments in order to measure the desired currents in these systems.

\subsection{Spin Hamiltonian and current operator}

Trapped-ion quantum simulation experiments enable controlled studies of interacting systems of spins~\cite{Blatt2012,Schneider2012,Monroe2021} as well as bosons~\cite{Porras2004a,Debnath2018}.
Though our scheme is general, we focus here on quantum simulation experiments for spin-$1/2$ degrees of freedom, which in trapped ions can be represented by two internal electronic states.
By coupling to the collective vibrational modes of the ion crystal, it is possible to engineer generic spin Hamiltonians of Heisenberg type~\cite{Porras2004}, in particular also those with isotropic spin--spin interaction in $x$- and $y$-direction~\cite{Jurcevic2014,Maier2019}.
The corresponding Hamiltonian is given by
\begin{equation}
	\label{eq:hamiltonianSpin}
	\mathcal{H} = - \sum_{\ell \neq \ell^\prime} J_{\ell \ell^\prime} S_{\ell}^{+} S_{\ell^\prime}^{-} + \mathcal{V} .
\end{equation}
Here, $S_{\ell}^{\pm} = S_{\ell}^x \pm i S_{\ell}^y$ are the spin raising and lowering operators at site~$\ell$, defined in terms of the local spin-$1/2$ operators~$S_{\ell}^{\alpha}$ with $\alpha \in \set{x, y, z}$, and $J_{\ell \ell^\prime} = J_{\ell^\prime \ell}^*$ are the (possibly complex~\cite{Manovitz2020}) interaction constants in $x$\nobreakdash- and $y$-direction. Furthermore, the term~$\mathcal{V}$ represents a possible spin--spin interaction in $z$-direction, which can be engineered using additional phononic modes~\cite{Porras2004}.
Important special cases of the Hamiltonian~\eqref{eq:hamiltonianSpin} include the $\mathrm{XY}$ model ($J_{\ell \ell^\prime} = J_{\ell \ell^\prime}^*$ and $\mathcal{V} = 0$) or the $\mathrm{XXZ}$ model ($J_{\ell \ell^\prime} = J_{\ell \ell^\prime}^*$ and $\mathcal{V} = - \sum_{\ell \ell^\prime} J_{\ell \ell^\prime}^z S_{\ell}^{z} S_{\ell^\prime}^{z}$), both of which are ubiquitous in many areas of physics and constitute paradigm models for strongly correlated materials~\cite{Giamarchi2004,Majlis2007,Sachdev2011}.
By virtue of the Holstein--Primakoff transformation~\cite{Holstein1940}, this Hamiltonian maps to the one in \cref{eq:hamiltonian} in the limit of hard-core bosons, by identifying the operators $S_\ell^+ = a_l$, $S_\ell^- = a_l^\dagger$, and $S_\ell^z = 1/2 - n_\ell$. In this mapping, the spin states $\ket{\uparrow}$ and $\ket{\downarrow}$ correspond to the bosonic vacuum $\ket{0}$ and the single excited state $\ket{1}$, respectively.

An important property of the Hamiltonian~\eqref{eq:hamiltonianSpin} is the conservation of the total magnetization in $z$-direction, giving rise to the local continuity equation
\begin{equation}
	\label{eq:continuity}
	\frac{\diff}{\diff t} S_{\ell}^{z} + \sum_{\ell^\prime \neq \ell} j_{\ell \ell^\prime} = 0 ,
\end{equation}
where $j_{\ell \ell^\prime}$ is the spin current operator from site $\ell$ to $\ell^\prime$.
As in the case of soft-core bosons discussed in the main text, the form of the current operator can be derived by comparing \cref{eq:continuity} to the Heisenberg equation of motion
\begin{equation}
\frac{\diff}{\diff t} S_{\ell}^{z} = -i \Commutator{S_{\ell}^{z}}{\mathcal{H}} .
\end{equation}
Using the commutation relations
\begin{subequations}
	\label{eq:commutationrelationsSpin}
\begin{align}
	\Commutator{S_\ell^z}{S_{\ell^\prime}^{\pm}} &= \pm \delta_{\ell \ell^\prime} S_{\ell}^{\pm} , \\
	\Commutator{S_{\ell}^{+}}{S_{\ell^\prime}^{-}} &= 2 \delta_{\ell \ell^\prime} S_{\ell}^{z} ,
\end{align}
\end{subequations}
we find
\begin{equation}
	\label{eq:currentoperatortrappedions}
	j_{\ell \ell^\prime} = -i \left( J_{\ell \ell^\prime} S_\ell^+ S_{\ell^\prime}^- - J_{\ell \ell^\prime}^* S_{\ell^\prime}^+ S_{\ell}^- \right) ,
\end{equation}
in complete analogy to \cref{eq:currentoperator}.

\subsection{\label{app:trappedions:protocol}Measurement protocol}

To implement the current measurement scheme in a trapped-ion system, we propose using collective vibrational modes of the ion crystal as ancillas.
Here, we consider the case of a single ancilla corresponding to a certain mode of an orthogonal set of phonon modes, e.g., the center-of-mass mode.
As before, we represent the ancilla by the bosonic annihilation and creation operators $b$ and $b^\dagger$.
Through an appropriate choice of the laser detunings, the ions can be coupled to this specific mode via the red sideband Hamiltonian~\cite{Wineland1998}
\begin{equation}
	\label{eq:hamiltonianredsideband}
	\mathcal{H}_\mathrm{cpl} = \frac{1}{2} \sum_{\ell} \Omega_\ell^{\mathrm{R}} \eta_\ell \left( S_\ell^+ b \etothepowerof{-i \varphi_\ell} + S_\ell^- b^\dagger \etothepowerof{i \varphi_\ell} \right) .
\end{equation}
Here, $\Omega_\ell^{\mathrm{R}}$ is the (Raman) Rabi frequency, $\eta_\ell$ is the Lamb--Dicke parameter, and $\varphi_\ell$ is the phase of the coupling to the $\ell$\nobreakdash-th ion, respectively.
This Hamiltonian has the same form as the general coupling Hamiltonian in \cref{eq:hamiltoniancouplinggeneral} with $A = \sum_\ell \lambda_\ell S_\ell^-$ and $\lambda_\ell = \Omega_\ell^{\mathrm{R}} \eta_\ell \etothepowerof{i \varphi_\ell} / 2 \Omega$, where $\Omega$ is the overall strength of the coupling pulse in \cref{eq:hamiltoniancouplinggeneral} ($M = 1$).

For an ancilla that is initially in its motional ground state, the results in \cref{eq:probabilitiesgeneral} immediately carry over to the trapped-ion case. 
Such a situation can be achieved thanks to the efficient cooling of trapped-ion phonon modes~\cite{Wineland1998,Eschner2003,Lechner2016,Feng2020,Chen2020}.
In addition, phonon heating is typically much slower than the coherent coupling pulses we are interested in here~\cite{Schindler2013}.
Nevertheless, in practice it may be desirable to relax the requirement of cooling the relevant phonon modes exactly to their motional ground states. Therefore, we consider here the more general case of an ancilla that is initially in the mixed state
\begin{equation}
	\rho_\mathrm{A} = \sum_{n = 0}^{\infty} p_n \ketbra{n}{n} ,
\end{equation}
where $p_n$ is the occupation probability of the $n$-th excited phonon state.
The precise distribution~$p_n$ is unimportant for the following discussion, but we require it to be quasistationary within the coupling duration~$\Delta t$.
This includes the common scenario of a thermal state, i.e., $p_n \equiv p_n(T) = \etothepowerof{-n \omega / T} / Z$, where $\omega$ is the frequency of the ancillary (center-of-mass) mode and $Z = (1 - e^{- \omega / T})^{-1}$ is the partition sum in a canonical ensemble at temperature~$T$ (here and in what follows, we set $\kb = 1$).

As before, we assume that when the coupling is turned on, the total state of system plus ancilla is given by a product state, $\rho = \rho_0 \otimes \rho_\mathrm{A}$.
Proceeding in an analogous way as in \cref{app:derivation:general}, we find that, to leading order in the effective coupling~$s = (\Omega \Delta t)^2$, the probability of detecting $n$ phonons in the ancilla reads
\begin{equation}
	\label{eq:probabilitytrappedions}
\begin{split}
	P(n) = p_n &-s \left[ \left( n + 1 \right) p_n - n p_{n - 1} \right] \Braket{A^\dagger A} \\
	&-s \left[ n p_n - \left( n + 1 \right) p_{n + 1} \right] \Braket{A A^\dagger} .
\end{split}
\end{equation}
For a thermal state with $T \ll \omega$, implying $p_n \approx \delta_{n 0}$, we recover the result in \cref{eq:probability0AdaggerA}.
According to \cref{eq:probabilitytrappedions}, the expectation values~$\braket{A^\dagger A}$ and $\braket{A A^\dagger}$ can be extracted by measuring how the phonon distribution has changed after the coupling with respect to the original (thermal) distribution.
Counting the phonon population is a common tool in modern trapped-ion experiments~\cite{Leibfried1996,Roos2000,Gebert2016,Um2016,Ding2017}.

To access the current~\eqref{eq:currentoperatortrappedions} between two ions~$\ell_1$ and $\ell_2$, we choose only those couplings in \cref{eq:hamiltonianredsideband} corresponding to $\ell_1$ and $\ell_2$ different from zero, i.e.,
\begin{equation}
	\label{eq:couplingoperatorSpin}
	A = \etothepowerof{i \theta_{\ell_1}} S_{\ell_1}^{-} + \etothepowerof{i \theta_{\ell_2}} S_{\ell_2}^{-} .
\end{equation}
The required single-site addressing is typically available in state-of-the art trapped-ion quantum-simulation experiments~\cite{Smith2016,Maier2019}.
The measurement scheme then gives access to the general combination
\begin{equation}
\begin{split}
	A^\dagger A &= S_{\ell_1}^+ S_{\ell_1}^- + S_{\ell_2}^+ S_{\ell_2}^- \\
	&\phantom{{}=} + \etothepowerof{i (\theta_{\ell_2} - \theta_{\ell_1})} S_{\ell_1}^+ S_{\ell_2}^- + \etothepowerof{-i (\theta_{\ell_2} - \theta_{\ell_1})} S_{\ell_2}^+ S_{\ell_1}^- .
\end{split}
\end{equation}
By choosing the phases of the coupling such that ${\theta_{\ell_2} - \theta_{\ell_1}} = \phi_{\ell_1 \ell_2} - \pi / 2$, where $\phi_{\ell_1 \ell_2} = \arg(J_{\ell_1 \ell_2})$, and using $S_\ell^+ S_\ell^- = S_\ell^z + 1 / 2$, we obtain
\begin{equation}
	A^\dagger A = \doublestruck{1} + S_{\ell_1}^z + S_{\ell_2}^z + \frac{j_{\ell_1 \ell_2}}{|J_{\ell_1 \ell_2}|} .
\end{equation}
For the coupling operator~\eqref{eq:couplingoperatorSpin}, the commutation relations~\eqref{eq:commutationrelationsSpin} imply $\commutator{A}{A^\dagger} = -2 (S_{\ell_1}^z + S_{\ell_2}^z)$, from which we conclude
\begin{equation}
	A A^\dagger = \doublestruck{1} - S_{\ell_1}^z - S_{\ell_2}^z + \frac{j_{\ell_1 \ell_2}}{|J_{\ell_1 \ell_2}|} .
\end{equation}
Inserting these expressions into \cref{eq:probabilitytrappedions}, we arrive at
\begin{equation}
	p_{\ell_1 \ell_2}(n) = p_n - s \left[ \alpha_n \left( 1 + \frac{\Braket{j_{\ell_1 \ell_2}}}{|J_{\ell_1 \ell_2}|} \right) + \beta_n \Braket{S_{\ell_1}^z + S_{\ell_2}^z} \right]
\end{equation}
with $\alpha_n = (2 n + 1) p_n - n p_{n - 1} - (n + 1) p_{n + 1}$ and $\beta_n = p_n - n p_{n - 1} + (n + 1) p_{n + 1}$.
As discussed in \cref{app:derivation:currents}, the quantity~$\braket{S_{\ell_1}^z + S_{\ell_2}^z}$ required to isolate the desired current~$\braket{j_{\ell_1 \ell_2}}$ can be obtained in a separate standard measurement, or its contribution can be eliminated by considering the anti-symmetric combination
\begin{equation}
	\frac{p_{\ell_1 \ell_2} - p_{\ell_2 \ell_1}}{2}(n) = p_n - s \alpha_n \frac{\Braket{j_{\ell_1 \ell_2}}}{|J_{\ell_1 \ell_2}|} .
\end{equation}

As these discussions show, the method works for general initial mixed states that are diagonal in the occupation basis of the ancilla. This feature may even be exploited to optimize the obtained signal.
The proposed trapped-ion implementation of our measurement scheme can immediately be extended to global currents, current variances, as well as current--current correlations, following similar ideas as presented in \cref{app:derivation}.

\bibliography{references}

\end{document}